\documentclass[10pt,journal,doublecolumn,twoside]{IEEEtran}

\usepackage{bbm}
\usepackage{amsfonts}
\usepackage{}
\IEEEoverridecommandlockouts

\usepackage{subfigure}
\usepackage{colortbl}
\usepackage{soul}
\soulregister\ref7 
\soulregister\cite7 
\soulregister\eqref7 
\usepackage{bm}

\usepackage{graphicx}  
\usepackage{url}       
\usepackage{rotating}
\usepackage{amsmath}   
\usepackage{cite}

\usepackage{amsfonts,amssymb}

\usepackage{stfloats}
\usepackage{exscale}
\usepackage{relsize}

\usepackage{cases}
\usepackage{algorithm}
\usepackage{multirow}
\usepackage{algorithmic}
\usepackage{amsthm}
\usepackage[utf8]{inputenc}

\newcommand{\myvec}[1]%
{\stackrel{\raisebox{-2pt}[0pt][0pt]
{\small$\rightharpoonup$}}{#1}}

\newcommand{\ls}[1]
    {\dimen0=\fontdimen6\the\font
     \lineskip=#1\dimen0
     \advance\lineskip.5\fontdimen5\the\font
     \advance\lineskip-\dimen0
     \lineskiplimit=.9\lineskip
     \baselineskip=\lineskip
     \advance\baselineskip\dimen0
     \normallineskip\lineskip
     \normallineskiplimit\lineskiplimit
     \normalbaselineskip\baselineskip
     \ignorespaces
    }

\usepackage{etoolbox}
\makeatletter
\patchcmd{\@makecaption}
  {\scshape}
  {}
  {}
  {}
\makeatother

\usepackage[absolute,showboxes]{textpos}

\TPMargin{3pt}

\begin{document}

\newcommand{\copyrightstatement}{
    \begin{textblock}{0.84}(0.08,0.95) 
         \noindent
         \footnotesize
         \copyright 2022 IEEE. Personal use of this material is permitted. Permission from IEEE must be obtained for all other uses, in any current or future media, including reprinting/republishing this material for advertising or promotional purposes, creating new collective works, for resale or redistribution to servers or lists, or reuse of any copyrighted component of this work in other works. DOI: 10.1109/JSYST.2021.3132612.
    \end{textblock}
}
\copyrightstatement
\setlength{\columnsep}{0.24in}

\title{Index Modulation Embedded Mode Hopping for Anti-Jamming}

\author{Liping Liang, \textit{Student Member, IEEE}, Wenchi Cheng, \textit{Senior Member, IEEE}, \\
Wei Zhang, \textit{Fellow, IEEE}, and Hailin Zhang, \textit{Member, IEEE}


\thanks{Part of this work has been presented in the IEEE Global Communications Conference, 2019~\cite{9013754}.
This work was supported by the National Key R\&D Program of China (2021YFC3002100), the National Natural Science Foundation of China (No. 61771368), Key Area Research and Development Program of Guangdong Province under grant No. 2020B0101110003, in part by the Australian Research Council's Project funding scheme under LP160101244, and in part by Shenzhen Science \& Innovation Fund under Grant JCYJ20180507182451820.

Liping Liang, Wenchi Cheng, and Hailin Zhang are with the State Key Laboratory of Integrated Services Networks, Xidian University, Xi'an, 710071, China (e-mails: liangliping@xidian.edu.cn; wccheng@xidian.edu.cn; hlzhang@xidian.edu.cn).

W. Zhang is with the School of Electrical Engineering and Telecommunications, The University of New South Wales, Sydney, Australia
(e-mail: w.zhang@unsw.edu.au).

}
}

\maketitle
\thispagestyle{empty}
\pagestyle{empty}

\begin{abstract}
Due to the crowded spectrum, frequency hopping (FH) techniques are now very difficult to achieve efficient anti-jamming and increase spectrum efficiency (SE) for wireless communications. The emerging orbital angular momentum (OAM), which is a property describing the helical phase fronts of electromagnetic waves, offers the potential to improve reliability and increase SE in wireless communications. To achieve efficient anti-jamming and increase SE of wireless communications with slight computational complexity cost, in this paper we propose an index-modulation embedded mode-hopping (IM-MH) scheme, which simultaneously activates several OAM-modes for hopping along with additional index information and signal information transmission. We analyze the average bit error rates (ABERs) for our proposed IM-MH scheme with perfect channel state information (CSI) and imperfect CSI, respectively. We also propose the index-modulation embedded double-serial MH (IM-DSMH) scheme, which randomly activates one OAM-mode as the serial second hop to transmit the hopping signals in the IM-MH scheme, to further decrease the ABER of wireless communications. Extensive numerical results demonstrate that our proposed schemes within a narrowband can achieve the low ABER and significantly increase the SE. Also, the ABERs of our proposed IM-MH and IM-DSMH schemes are around 25\% and 10\%, respectively, compared with that of the mode hopping scheme.

\end{abstract}


\section{Introduction}~\label{sec:Intro}

\IEEEPARstart{F}{requency} hopping (FH) has been widely used for secure communications under hostile jamming. In conventional FH communications, the carrier frequency is quickly changed according to a preset hopping pattern, which facilities the avoidance of jamming attacks. There are several typical FH schemes, such as uncoordinated FH (UFH), message-driven FH (MDFH), and adaptive FH (AFH)~\cite{2010_UFH,2012_UFH,xiao,2016_AFH}. To overcome the dependence of preset hopping patterns and guarantee efficient anti-jamming of highly dynamic wireless networks, the authors proposed the UFH scheme, which randomly selects one channel to hop at both transmitter and receiver without any predefined hopping patterns~\cite{2010_UFH,2012_UFH}. Compared with the conventional FH schemes, UFH is with lower spectrum efficiency (SE) due to the lack of coordination between the transmitter and receiver. In MDFH, the transmit information contains encryption information and signal information, thus increasing the SE of FH communications. AFH schemes attempt to adaptively select a channel with minimum interference~\cite{2016_AFH}.
However, as the traffic date rate greatly increases and serve diversity explodes, the limited bandwidth results in the spectrum being more and more crowded in wireless communications, thus increasing the difficulty of anti-jamming with FH. Also, the scarce spectrum limits the SE of FH schemes.

Orbital angular momentum (OAM), which describes the helical phase fronts of electromagnetic waves, offers a new degree of freedom different from the conventional degrees of freedom such as time, frequency, and polarization~\cite{OAM_mag,7968418,7797488,9530523,OAM_MIMO_Ren,OEM}. By using the orthogonal and independent characteristic of integer OAM-modes, beams carrying different orders of OAM-modes can simultaneously propagate on a narrowband with less or no inter-mode interference, which shows potential applications of OAM for efficient multiplexing~\cite{OAM_mag,7968418,7797488,9530523,OAM_MIMO_Ren,OEM} and anti-jamming~\cite{Mirhosseini:13,Willner:15,secure_OAM,8712342,MH}. In the past few decades years, OAM of light has been studied for data encoding and channel hopping to increase the tolerance to hostile jamming attacks in optical communications and quantum communications. To rapidly switch the generated OAM-modes, a digital micro-mirror device~\cite{Mirhosseini:13} and a programable spatial light modulator~\cite{Willner:15} were used to change the hologram in free-space, which only generated one OAM-mode at each time slot. Due to the inefficient OAM-mode utilization, these schemes are with very low SE.

\begin{figure*}
\centering
  \includegraphics[width=0.95\textwidth]{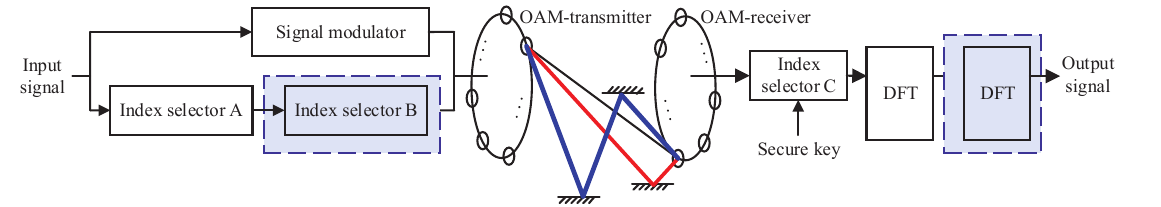}\\
  \caption{The system model of OAM-based index modulation wireless communications.}\label{fig:sys}
\vspace{-5pt}
\end{figure*}

By contrast to optical communications, there are few researches about the anti-jamming applications of OAM in wireless communications~\cite{9306012,reliable_encrypt,secure_OAM,8712342,MH}. To avoid the information interception by eavesdroppers, the authors of~\cite{8712342} proposed a secure rang-dependent transmission scheme, where the OAM-modes were generated by frequency diverse array antennas. Mode-hopping (MH) scheme, which rapidly switches the operating OAM-mode to another according to a certain MH pattern controlled by pseudo-noise generators within a narrowband, has been studied to achieve robust anti-jamming of wireless communications~\cite{MH}. However, the main limitation of this technique is that it only transmits legitimate signal information over one OAM-mode, thus resulting in low SE. Introducing the concept of index modulation~\cite{MIMO_OFDM_IM}, the authors of~\cite{Basar_IM} have proposed the OAM with index modulation (OAM-IM) scheme to decrease the bit error rate and increase the SE. In addition to transmit traditional signal constellations, the OAM-IM scheme also utilized the indices of OAM-modes for additional index information transmission. However, when multiple OAM-modes are activated, the probability that signals are jammed significantly increases under hostile jamming attack environments, thus resulting in inefficient anti-jamming results. How to use OAM within a limited spectrum band to increase SE and achieve efficient anti-jamming remains an open and interesting problem in wireless communications.

To overcome the problem mentioned above, in this paper we propose the index modulation embedded MH (IM-MH) scheme, which simultaneously activates serval OAM-modes for hopping in wireless communications. Different from the conference version~\cite{9013754}, where we mainly focus on deriving the optimal number of activated OAM-modes to achieve the maximum capacity, we extend the IM-MH scheme from one hop to multiple hops within a symbol duration, thus further improving the reliability of our proposed IM-MH scheme. We derive the average bit error rates (ABERs) of our proposed IM-MH scheme under perfect CSI and imperfect CSI scenarios, respectively, to evaluate the anti-jamming performance. The jamming probability of IM-MH scheme increases as the number of activated OAM-modes increases at each time-slot, thus resulting in the relatively high ABER. To further decrease the ABER, we also propose the index modulation embedded double-serial MH (IM-DSMH) scheme, which randomly activates one OAM-mode as the serial second hop to transmit the hopping signals in the IM-MH scheme. We conduct extensive numerical results to evaluate our proposed schemes, showing that our proposed schemes within a narrowband can achieve efficient anti-jamming and high SEs in wireless communications.

The remainder of this paper is organized as follows. Section~\ref{sec:sys} gives the OAM-based index modulation wireless communications system model. To reduce the ABER and increase the SE, the IM-MH scheme is proposed in Section~\ref{sec:MH}, where we derive the exact closed-form expressions of ABER under perfect and imperfect CSI scenarios, respectively. The IM-DSMH scheme is proposed to further decrease the ABER of wireless communications in Section~\ref{sec:DMMH}. Numerical results of proposed schemes are discussed in Section~\ref{sec:simu}. The paper concludes with Section~\ref{sec:conc}.


\vspace{-5pt}

\section{System Model}~\label{sec:sys}
We build up a new OAM-based index modulation wireless communications system model in Fig.~\ref{fig:sys}, which consists of a signal modulator, an index selector A, and an OAM-transmitter at the transmitter followed by an OAM-receiver, an index selector C, and a discrete Fourier transform (DFT) operator at the receiver. Index selectors A and C are synchronized. When index selector B at the transmitter as well as another DFT operator at the receiver are added into the system, the DSMH can work for the OAM-based index modulation wireless communications system. Index selector C is synchronized with index selectors A and B when the DSMH works.

As shown in Fig.~\ref{fig:sys}, the input signal information is separated to signal information and index information~\cite{Basar_IM}. According to the index information determined by secure keys, index selector A activates $I$ out of $N$ OAM-modes so that OAM multiplexing can be achieved at each hop. 
The remaining $(N-I)$ OAM-modes are inactivated. When the DSMH works, based on the index information one OAM-mode is activated by index selector B as the second serial hop. The OAM-modes are activated regularly for the transmitter and the receiver, but randomly for attackers. OAM-transmitter, which consists of $N$ elements equidistantly distributed around the circle of uniform circular array (UCA) antenna, can generate $N$ OAM-modes in wireless communications~\cite{OAM_imaging}. OAM-receiver is also comprised of $N$ elements equidistantly distributed around the circle of UCA antenna. OAM-transmitter and OAM-receiver can be coaxial and non-coaxial in the OAM-based index modulation wireless communications system. To simplify analysis below, we assume the OAM-transmitter and OAM-receiver are coaxial and aligned. At the receiver, index selector C is used to select the same OAM-modes activated by the transmitter. When the IM-MH works, the DFT operator is used to decompose OAM signals. Since one OAM-mode is activated by index selector B as the second serial hop, the basedband signals are modulated twice in the angular domain when the DSMH works. Therefore, two DFT operators are used to successively decompose OAM signals. In this paper, we adopt the fast hopping method with $U$ hops during a symbol transmission, which brings the diversity gain.

\section{The IM-MH Scheme}~\label{sec:MH}
In this section, we propose the IM-MH scheme to increase the SE while achieving the low ABER of wireless communications. Specifically, we first give the transmit signals and de-hop the corresponding received signals. Then, we derive the exact closed-form expression of ABER under the perfect CSI scenarios for arbitrary number of activated OAM-modes and hops. Finally, we calculate the ABER under the realistic imperfect CSI scenarios.

\subsection{Signal Transmission}
\begin{figure}
\centering
  \includegraphics[width=0.25\textwidth]{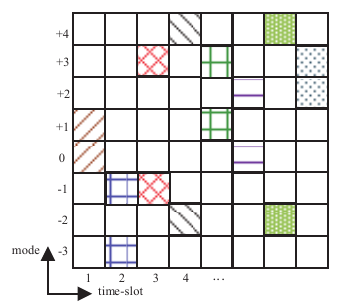}\\
  \caption{An example of IM-MH pattern. The time-mode resource blocks with specified color denote the activated OAM-modes at each hop. The activated OAM-modes are quickly changed according to the preset IM-MH pattern.}\label{fig:IM_MH}
\vspace{-5pt}
\end{figure}

Since each OAM combination corresponds to the specific index information, the number of available OAM combinations is $K_{1}=2^{\left\lfloor \log_{2}\binom{N}{I}\right\rfloor}$ when $I$ out of $N$ OAM-modes are simultaneously activated at each hop, where $\lfloor\cdot\rfloor$ is the floor function. We select a unique combination out of $K_{1}$ at each hop. To illustrate the hopping clearly, an example of IM-MH pattern is shown in Fig.~\ref{fig:IM_MH}, where the axis of OAM-mode is ranked in the ascending order of OAM-modes and we set $N=8$ as well as $I=2$. OAM-mode and time slot are integrated into a two-dimension time-mode resource block. The time-mode resource blocks with specified color denote the activated OAM-modes at each hop in Fig.~\ref{fig:IM_MH}. For instance, the coordinates $(1,0)$ and $(1,+1)$ represent that OAM-modes 0 and +1 are activated for hopping at the first time slot.

$I$ OAM-modes are activated to convey the index information and each activated OAM-mode conveys $M$-ary constellation symbols. Thus, the total number of transmission bits, denoted by $\eta_{0}$, corresponding to $U$ hops is obtained as follows:
 \begin{equation}
   \eta_{0}=\underbrace{I\log_{2}M}_{\eta_{s}} + \underbrace{U \log_{2}K_{1}}_{\eta_{x}},
   \label{eq:eta}
\end{equation}
where $\eta_{s}$ and $\eta_{x}$ are the signal transmission bits and the index information bits, respectively. 

Without loss of generality, we consider the $u$-th ($1 \leq u \leq U$) hop. According to the index information, the set of activated OAM-modes for the $u$-th hop, which is ranked in the ascending order of OAM-modes, is given by
\begin{equation}
    L_{u}=\{l_{1,u},\cdots, l_{i,u}, \cdots, l_{I,u}\},
\end{equation}
where $l_{i,u} (1 \leq i \leq I$ and $ |l_{i,u}|\leq N/2)$ is the $i$-th activated OAM-mode of all $I$ OAM-modes.

We denote by $\mathcal{S}$ the whole symbol constellations. According to the signal information, the transmitted modulated symbols, denoted by $\bm{s}_{u}$, at the output of $M$-ary constellation modulator is given by
\begin{equation}
   \bm{s}_{u}=[s_{1,u} \cdots s_{i,u}\cdots s_{I,u} ]^{T},
\end{equation}
where $[\cdot]^T$ represents the transpose of a vector and $s_{i,u} \in \mathcal{S}$. It is noticed that the modulated symbols are transmitted within $U$ hops duration. The emitted modulated signal, denoted by $x_{n,u}$, for the $n$-th $(0 \leq n \leq N-1)$ transmit element is expressed as follows:
\begin{eqnarray}
    x_{n,u}=\sum_{i=1}^{I}{s}_{i,u} e^{j\frac{2\pi n}{N} l_{i,u}}.
\end{eqnarray}

Emitted signals can be propagated in a line-of-sight (LoS) path~\cite{OAM_MIMO_Ren} or in sparse multipath environments containing an LoS path and several non-line-of-sight (NLoS) paths~\cite{HODM_2019,OAM_OFDM_ch}. The NLoS paths consist of several primary reflection paths, secondary reflection paths, and triple reflection paths in sparse multipath environments~\cite{Goldsmith_2005}.

Then, the received signal, denoted by $y_{m,u}$, for the $m$-th ($0 \leq m \leq N-1$) receive element is obtained as follows:
\begin{eqnarray}
    y_{m,u}=\sum_{n=0}^{N-1}h_{mn}x_{n,u}+ J_{m} + n_{m},
     \label{eq:y_mk}
\end{eqnarray}
where $h_{mn}$ represents the channel gain from the $n$-th legitimate transmit element to the $m$-th legitimate receive element, $J_{m}$ denotes the received jamming signals for the $m$-th legitimate receive element, and $n_{m}$ represents the received complex Gaussian noise with zero mean for the $m$-th legitimate receive element.

The secure key is pre-shared between the transmitter and receiver. We assume that the transmitter and receiver are completely synchronized. Thus, index selector C can activate the OAM-modes same with those activated by index selector A. To obtain de-hopping signals, the DFT operation is performed with respect to OAM-mode $ l_{i,u}$. Therefore, we have the de-hopping signal, denoted by $y_{{i,u}}$, corresponding to the OAM-mode $l_{i,u}$ as follows:
\begin{eqnarray}
    y_{{i,u}}=\frac{1}{N}\sum_{m=0}^{N-1} y_{m,u} e^{-j\frac{2\pi m}{N} l_{i,u}}. \label{eq:y_i_k_1}
\end{eqnarray}

In sparse multipath environments, the received interference contains the inter-mode interference, the inter-symbol interference, and the jamming signals. The inter-mode interference and inter-symbol interference, which are caused by the phase difference between the complex channel gains of LoS path and all reflection paths, can be mitigated by using channel ray-tracing and phase difference compensation methods~\cite{HODM_2019}.
Note that for the case of non-coaxis between the OAM-transmitter and OAM-receiver, beam steering or phase compensation can mitigate the inter-mode interference caused by the misalignment of transceiver~\cite{HODM_2019,8269275}. Since OAM is very sensitive to azimuthal angles, the small phase difference between the azimuthal angles of attackers and legitimate transceivers will lead to jamming failure. Therefore, any attacker attempting to interfere with legitimate transceiver requires the same OAM-modes and azimuthal angles. To simplify the analyses of ABER, we assume that attackers randomly activate $I$ out of $N$ OAM-modes at each hop without knowing the legitimate transceiver's IM-MH pattern. Also, attackers have the same azimuthal angles with legitimate transceiver. Thereby, the jamming signals can be mitigated with the help of DFT algorithm. The legitimate signals only can be jammed by the jamming signals with OAM-mode $l_{i,u}$ and the cross-talk of jamming signals on OAM-mode $l_{i,u}$.

It is noticed that index selector C is necessary. If index selector C is not used at the receiver, all OAM signals regardless of whether OAM-modes are activated or inactivated are decomposed simultaneously after DFT algorithm in IM-MH communications. Thus, the receiver will decompose unexpected OAM signals if attackers transmit signals with inactivated OAM-modes. For example, we have $L_{u}=\{-1, 0, +2\}$ at the transmitter. However, the orders of OAM-modes of decomposed signals are corresponding to $\{-1,0,+1,+2\}$ at the receiver, which implies that the decomposed signal corresponding to OAM-mode +1 is the jamming signal.

Therefore, the de-hopping signal $y_{i,u}$ in Eq.~\eqref{eq:y_i_k_1} is obtained as follows:
\begin{eqnarray}
   y_{i,u}=h_{l_{i,u}}s_{i,u}+\kappa_{i,u} w_{i,u}+n_{i,u},
   \label{eq:yi}
 \end{eqnarray}
where $h_{l_{i,u}}$, $w_{i,u}$, and $n_{i,u}$ represent the channel gain from the legitimate user's OAM-transmitter to the OAM-receiver, the interference from attackers, and the received noise with zero mean and variance $\sigma^{2}$, respectively, corresponding to the OAM-mode $l_{i,u}$. We assume that the interference follows Gaussian distribution with zero mean and variance $\sigma_{J}^{2}$. $\kappa_{i,u}$ is used to identify the existence of interference. When $\kappa_{i,u}=0$, it means that there is no interference in OAM-mode $l_{i,u}$. On the other hand, when $\kappa_{i,u}=1$, the legitimate signals are jammed by jamming signals.

  \setcounter{equation}{12}
\begin{figure*}[ht]
    \begin{eqnarray}
    P_{I,U}(\bm{s}\!\rightarrow \!\hat{\bm{s}}|\bm{H}_{u},I_{u}^{\prime})\!\!\!\!\!&=&\!\!\!\!\!\!\Pr \!\!\left[ \sum_{u=1}^{U}\!\!\left(\!\!\sum\limits_{i=1}^{I_{u}^{\prime}}\!\!\frac{|y_{i,u}\!\!-\!\!h_{l_{i,u}}\hat{s}_{i,u}|^{2}}{\sigma_{J}^{2}+\sigma^{2}}
\!\! +\!\!\!\!\!\!\sum\limits_{i=1+I_{u}^{\prime}}^{I}\!\!\frac{|y_{i,u}\!-\!h_{l_{i,u}}\hat{s}_{i,u}|^{2}}{\sigma^{2}}\!\!\right)
 \!\!< \!\!\sum_{u=1}^{U}\!\!\left(\!\!\sum\limits_{i=1}^{I_{u}^{\prime}}\!\!\frac{|y_{i,u}\!\!-\!\!h_{l_{i,u}}s_{i,u}|^{2}}{\sigma_{J}^{2}+\sigma^{2}}
 \!\!+\!\!\!\!\sum\limits_{i=1+I_{u}^{\prime}}^{I}\!\!\!\!\frac{|y_{i,u}\!-\!h_{l_{i,u}}s_{i,u}|^{2}}{\sigma^{2}}\!\!\right)\!\!\right]
 \nonumber\\
 &=& \!\!\!\!\!\! \Pr (D>0),
    \label{eq:PEP_Ge}
\end{eqnarray}
\hrulefill
\vspace{-5pt}
\end{figure*}

\setcounter{equation}{15}
\begin{figure*}[ht]
    \begin{equation}
    P_{I,U}(\bm{s} \!\rightarrow \!\hat{\bm{s}}|\bm{H}_{u},I_{u}^{\prime})
    \!=\!Q\left(\sqrt{\sum_{u=1}^{U}\left(\frac{ \sum\limits_{i=1}^{I_{u}^{\prime}} |h_{l_{i,u}}\Delta_{i,u}|^{2}}{2(\sigma^{2}+\sigma_{J}^{2})}
    +\frac{\sum\limits_{i=1+I_{u}^{\prime}}^{I}|h_{l_{i,u}}\Delta_{i,u}|^{2}}{2\sigma^{2}}\right)}\right),
   \label{eq:PEP_2}
\end{equation}
\hrulefill
\vspace{-5pt}
\end{figure*}

In sparse multipath environments, the effect of a dominant signal arriving with several weaker NLoS paths gives rise to Rician distribution. For the given transceiver, the channel of the LoS path, denoted by $h_{l_{i,u},LoS}$, corresponding to OAM-mode $l_{i,u}$ can be derived as follows~\cite{HODM_2019}:
    \begin{equation}
    \setcounter{equation}{8}
h_{l_{i,u},LoS}\!\!=\!\!\frac{\beta \lambda N j^{-l_{i,u}}e^{-j\frac{2\pi}{\lambda}\! \sqrt{d^{2}+r_{1}^{2}+r_{2}^{2}}}}{4\pi \sqrt{d^{2}+r_{1}^{2}\!+\!r_{2}^{2}}} \! J_{l_{i,u}}\!\!\!\left(\!\!\frac{2\pi r_{1} r_{2}}{\lambda \sqrt{d^{2}\!+\!r_{1}^{2}\!+\!r_{2}^{2}}} \!\!\!\right),
\label{eq:channel}
\end{equation}
where $\beta$ represents attenuation, $\lambda$ denotes the carrier wavelength, $d$ is the distance from the legitimate user's OAM-transmitter center to the OAM-receiver center for the LoS path, $r_{1}$ denotes the radius of legitimate user's OAM-transmitter, $r_{2}$ represents the radius of legitimate user's OAM-receiver, and $J_{l_{i,u}}(\cdot)$ is the $l_{i,u}$-th order of first kind Bessel function. Also, we denote by $h_{l_{i,u},NLoS}$ the channel of NLoS paths, which follows Rayleigh distribution with zero mean and variance $\sigma_{i,u}^{2}$. Therefore, we have the expression of $h_{l_{i,u}}$ as follows~\cite{fading_2005}:
 \begin{eqnarray}
    h_{l_{i,u}}=\sqrt{\frac{\xi}{1+\xi}}h_{l_{i,u},LoS}+\sqrt{\frac{1}{1+\xi}}h_{l_{i,u},NLoS}, \label{eq:h_ik}
   \end{eqnarray}
where $\xi$ called Ricain factor is the ratio between the power in the LoS path and the power in NLoS paths. 

\subsection{ABER Analysis Under Perfect CSI}~\label{subsec:PEP}
In this subsection, we analyze the ABER under perfect CSI at the receiver. We assume that $I_{u}^{\prime}\ (0 \leq I_{u}^{\prime} \leq I)$ out of $I$ OAM-modes are jammed at the $u$-th hop. We denote by $\bm{y}_{u}=[y_{1,u} \cdots y_{i,u} \cdots y_{I,u}]^{T}$ the received signal vector and ${\bm{H}}_{u}$ the diagonal channel matrix with entries $h_{l_{i,u}}$, respectively, for the $u$-th hop. Based on Eq.~\eqref{eq:yi}, the conditional probability density function (PDF), denoted by $f(\bm{y}_{u}|\bm{H}_{u}, I_{u}^{\prime})$, of $\bm{y}_{u}$ with $I_{u}^{\prime}$ OAM-modes jammed under perfect CSI is given as follows:
\begin{eqnarray}
 &&\hspace{-1.2cm}f(\bm{y}_{u}|\bm{H}_{u}, I_{u}^{\prime})
 \nonumber\\
 &&\hspace{-1.2cm}=\frac{\exp \!\!\left(-\sum\limits_{i=1}^{I_{u}^{\prime}}\frac{|y_{i,u}-h_{i,u}s_{i,u}|^{2}}{\sigma_{J}^{2}+\sigma^{2}}\!
 -\!\!\sum\limits_{i=1+I_{u}^{\prime}}^{I}\frac{|y_{i,u}-h_{i,u}s_{i,u}|^{2}}{\sigma^{2}}\right)}{\pi^{I}(\sigma_{J}^{2}+\sigma^{2})^{I_{u}^{\prime}}
 \sigma^{2(I-I_{u}^{\prime})}}.
\end{eqnarray}

Assuming that $U^{\prime}$ ($0 \leq U^{\prime}  \leq U$) out of $U$ hops are jammed, we have the conditional PDF, denoted by $f(\bm{y}_{1}\cdots \bm{y}_{U} |\bm{H}_{1}\cdots \bm{H}_{U}, I_{1}^{\prime}\cdots I_{U}^{\prime})$, corresponding to $U$ hops as follows:
\begin{equation}
 f(\bm{y}_{1}\cdots \bm{y}_{U} |\bm{H}_{1}\cdots \bm{H}_{U}, I_{1}^{\prime}\cdots I_{U}^{\prime})=\prod_{u=1}^{U} f(\bm{y}_{u}|\bm{H}_{u}, I_{u}^{\prime}).
\end{equation}

The transmitter and the receiver per-share the MH pattern under the control of secret keys. Based on the pre-shared MH pattern and synchronization, the receiver knows which OAM-modes are activated by the transmitter. We use the maximum likelihood (ML) decoder at the receiver to estimate the bits related to transmit symbols. Searching all signal constellations, we have
\begin{eqnarray}
    \hat{\bm{s}} \hspace{-0.3cm}&=&\hspace{-0.3cm} \arg \max_{\bm{s}\in \mathcal {S}} \ \ln f(\bm{y}_{1}\cdots \bm{y}_{U} |\bm{H}_{1}\cdots \bm{H}_{U}, I_{1}^{\prime}\cdots I_{U}^{\prime})
    \nonumber\\
    &=& \hspace{-0.3cm}\arg \min_{\bm{s}\in \mathcal {S}}  \left\{\sum_{u=1}^{U}\left(\sum\limits_{i=1}^{I_{u}^{\prime}}\frac{|y_{i,u}-h_{l_{i,u}}s_{i,u}|^{2}}{\sigma_{J}^{2}+\sigma^{2}} \right.\right.
    \nonumber\\
 &&\hspace{1cm}\left.\left.+\sum\limits_{i=1+I_{u}^{\prime}}^{I}\frac{|y_{i,u}-h_{l_{i,u}}s_{i,u}|^{2}}{\sigma^{2}}\right)\right\},
  \label{eq:ml}
\end{eqnarray}
where $\bm{s}=[\bm{s}_{1}\cdots \bm{s}_{u}\cdots\bm{s}_{U}]$ and $\hat{\bm{s}}$ is the estimation of $\bm{s}$.

Based on Eq.~\eqref{eq:ml}, we can derive the expression of conditional pairwise error probability (PEP), denoted by $P_{I,U}(\bm{s} \!\rightarrow \!\hat{\bm{s}}|\bm{H}_{u},I_{u}^{\prime})$, of $\hat{\bm{s}}$ detected actually instead of $\bm{s}$ as shown in Eq.~\eqref{eq:PEP_Ge}, where $D$ is the difference between the two sides of unequal sign and $\hat{s}_{i,u}$ is the estimation of $s_{i,u}$.

For simplicity, we denote by $\Delta_{i,u}={s}_{i,u}-\hat{s}_{i,u}$. Since the received interference and noise follow complex Gaussian distribution, the mean and variance of $D$, denoted by $D_{\mu}$ and $D_{var}$, are derived as
\setcounter{equation}{13}
\begin{equation}
    D_{\mu}=-\sum_{u=1}^{U}\left(\frac{ \sum\limits_{i=1}^{I_{u}^{\prime}} |h_{l_{i,u}}\Delta_{i,u}|^{2}}{\sigma^{2}+\sigma_{J}^{2}}
    +\frac{\sum\limits_{i=1+I_{u}^{\prime}}^{I}|h_{l_{i,u}}\Delta_{i,u}|^{2}}{\sigma^{2}}\right)
\end{equation}
and
\begin{equation}
  D_{var}=2\sum_{u=1}^{U}\left(\frac{ \sum\limits_{i=1}^{I_{u}^{\prime}} |h_{l_{i,u}}\Delta_{i,u}|^{2}}{\sigma^{2}+\sigma_{J}^{2}}
    +\frac{\sum\limits_{i=1+I_{u}^{\prime}}^{I}|h_{l_{i,u}}\Delta_{i,u}|^{2}}{\sigma^{2}}\right),
\end{equation}
respectively. Thus, Eq.~\eqref{eq:PEP_Ge} can be re-expressed as Eq.~\eqref{eq:PEP_2}, where
\setcounter{equation}{16}
\begin{equation}
    Q(x)\approx\frac{1}{12}e^{-\frac{x^{2}}{2}}+\frac{1}{4}e^{-\frac{2x^{2}}{3}}
    \label{eq:Q_func}
\end{equation}
is the Gaussian Q-function~\cite{Q_function}. Observing Eq.~\eqref{eq:PEP_2}, the conditional PEP depends on the Euclidean distance between ${\hat{s}}_{i,u}$ and ${s}_{i,u}$ for given $\bm{H}_{u}$ and $I_{u}^{\prime}$.
Aiming at calculating the ABER to evaluate the anti-jamming performance of our proposed IM-MH scheme, we need to statistically average the conditional PEP over the number of OAM-modes jammed by interference, the number of hops jammed, and the channel gains. In the following, we obtain the exact closed-form expression of conditional PEP for arbitrary $U^{\prime}$ and $I_{u}^{\prime}$.

{\textit{Proposition 1}:} The exact conditional PEP, denoted by $P_{I,U}(\bm{s} \!\rightarrow \!\hat{\bm{s}}|I_{u}^{\prime})$, with $I_{u}^{\prime}$ OAM-modes jammed by hostile jamming under perfect CSI is derived as follows:
    \begin{eqnarray}
    P_{I,U}(\bm{s} \!\rightarrow \!\hat{\bm{s}}|I_{u}^{\prime})\!\!\!\!&=&\!\!\!\! \frac{\exp\left(\sum\limits_{i=1}^{I}\sum\limits_{u=1}^{U}\frac{\xi\rho_{i,u}{h}^{2}_{l_{i,u},LoS}}{4(1+\xi)-\rho_{i,u}\sigma_{i,u}^{2}{h}^{2}_{l_{i,u},LoS}}\right)}
   {12\prod\limits_{i=1}^{I}\prod\limits_{u=1}^{U}\left[1-\frac{\rho_{i,u}\sigma_{i,u}^{2}{h}^{2}_{l_{i,u},LoS}}{4(1+\xi)}\right]}
   \nonumber\\
 \!\!\!\! \!\!\!\!  \!\!\!\!   && \!\!\!\! + \frac{\exp\left(\sum\limits_{i=1}^{I}\sum\limits_{u=1}^{U}\frac{\xi\rho_{i,u}{h}^{2}_{l_{i,u},LoS}}{3(1+\xi)-\rho_{i,u}\sigma_{i,u}^{2}{h}^{2}_{l_{i,u},LoS}}\right)}
   {4\prod\limits_{i=1}^{I}\prod\limits_{u=1}^{U}\left[1-\frac{\rho_{i,u}\sigma_{i,u}^{2}{h}^{2}_{l_{i,u},LoS}}{3(1+\xi)}\right]},
   \nonumber\\
   \label{eq:PEP_U}
\end{eqnarray}
where $\rho_{i,u}$ is given by
\begin{equation}
    \rho_{i,u}=\left\{
    \begin{aligned}
       &-\frac{\Delta_{i,u}^{2}}{\sigma^{2}+\sigma_{J}^{2}}  \ \ \ \text{if}\ i\in [1,I_{u}^{\prime}];\\
       &-\frac{\Delta_{i,u}^{2}}{\sigma^{2}} \ \ \ \ \ \ \  \text{if}\ i\in [1+I_{u}^{\prime},I].
    \end{aligned}
    \right.
    \label{eq:rho}
\end{equation}
\begin{proof}
    See Appendix~\ref{app:P1}.
\end{proof}

To average the conditional PEP over $I_{u}^{\prime}$ and $U^{\prime}$, the jamming probability is required. We assume each OAM-mode has an equal probability for being active by using the equiprobable activation mapping method~\cite{7564470}. Thus, each OAM-mode is jammed with equal probability. We denote by $P(I_{u}^{\prime}|I)$ the jamming probability that $I_{u}^{\prime}$ out of $I$ OAM-modes are jammed at the $u$-th hop. Thus, we have
\begin{equation}
    P(I_{u}^{\prime}|I)=\frac{\binom {I} {I_{u}^{\prime}}\binom {N-I} {I-I_{u}^{\prime}}}{\binom {N} {I}}.
\label{eq:jamming_p}
\end{equation}
When $N-I < I-I_{u}^{\prime}$, we have $P(I_{u}^{\prime}|I)=0$. Also, the probability, denoted by $P_{0}$, without any interference at the $u$-th hop is $\binom {N-I} {I}/\binom {N} {I}$. Thus, we can calculate the jamming probability, denoted by $P(U^{\prime}|U)$, with $U^{\prime}$ out of $U$ hops jammed as follows:
\begin{eqnarray}
   \!\!\!\!\! P(U^{\prime}|U)\!\!\!\!&=&\!\!\!\!\underbrace{\sum\limits_{I_{1}^{\prime}=1}^{I}\sum\limits_{I_{2}^{\prime}=1}^{I}\cdots\sum\limits_{I_{U^{\prime}}^{\prime}=1}^{I}}
    _{U^{\prime}-fold}
    \binom {U} {U^{\prime}}P_{0}^{U-U^{\prime}}
    \nonumber\\
    &&\!\!\!\!\!\!\!\times \prod\limits_{I_{i}^{\prime}=1}^{I} \binom {I} {I_{i}^{\prime}}\underbrace{P(I_{1}^{\prime}|I)P(I_{2}^{\prime}|I)\cdots P(I_{U^{\prime}}^{\prime}|I)}_{U^{\prime}-{\rm fold}}.
    \label{eq:jp}
\end{eqnarray}

The upper bound on ABER can be calculated by the following steps. First, we weight conditional PEP $P(U^{\prime}|U)$ by the number of transmit bit errors related to a given transmission sequence. Next, we sum the conditional PEP over all error events corresponding to the given transmit sequence. Then, we statistically average the sum over all possible transmit sequences. Finally, we average the value over the number of transmission bits per symbol~\cite{Digital_C}. Therefore, based on Eqs.~\eqref{eq:PEP_U} and \eqref{eq:jp}, the ABER, denoted by $P_{I,U}$, is calculated using the union-bound method as follows:
\begin{equation}
   P_{I,U} \leq \sum\limits_{\bm{s}}^{2^{\eta_{s}}}
   \sum\limits_{\hat{\bm{s}}}^{2^{\eta_{s}}}\sum\limits_{U^{\prime}=0}^{U}\frac{P(U^{\prime}|U)P_{I,U}(\bm{s} \rightarrow \hat{\bm{s}}|I_{u}^{\prime})N_{e}(\bm{s}, \hat{\bm{s}})}{\eta_{0} 2^{\eta_{s}}},
   \label{eq:APEP}
\end{equation}
where $N_{e}(\bm{s}, \hat{\bm{s}})$ represents the number of transmission bit errors for the event ($\bm{s} \rightarrow \hat{\bm{s}}$).

\setcounter{equation}{25}
\begin{figure*}[ht]
\begin{small}
        \begin{eqnarray}
    P_{I}(\bm{s}\!\rightarrow \!\hat{\bm{s}}|\widetilde{\bm{H}}_{u},I_{u}^{\prime})\!\!\!\!\!&=&\!\!\!\!\!\Pr \!\!\left[ \! \sum_{u=1}^{U}\!\!\left(\!\!\sum\limits_{i=1}^{I_{u}^{\prime}}\!\!\frac{|y_{i,u}\!-\!\tilde{h}_{l_{i,u}}\hat{s}_{i,u}|^{2}}{\sigma_{J}^{2}\!+\!\sigma^{2}\!+\!\sigma_{\epsilon}^{2}
    |\hat{s}_{i,u}|^{2}}
\! +\!\!\!\!\sum\limits_{i=1+I_{u}^{\prime}}^{I}\!\!\!\!\frac{|y_{i,u}\!\!-\!\!\widetilde{h}_{l_{i,u}}\hat{s}_{i,u}|^{2}}{\sigma^{2}\!
+\!\sigma_{\epsilon}^{2}|\hat{s}_{i,u}|^{2}}\!\!\right)
\!\! < \!\! \sum_{u=1}^{U}\!\!\left(\!\!\sum\limits_{i=1}^{I_{u}^{\prime}}\!\!\frac{|y_{i,u}\!-\!\widetilde{h}_{l_{i,u}}s_{i,u}|^{2}}{\sigma_{J}^{2}\!+\!\sigma^{2}\!+\!\sigma_{\epsilon}
^{2}|s_{i,u}|^{2}}
 \!+\!\!\!\!\sum\limits_{i=1+I_{u}^{\prime}}^{I}\!\!\frac{|y_{i,u}\!-\!\widetilde{h}_{l_{i,u}}s_{i,u}|^{2}}{\sigma^{2}\!+\!\sigma_{\epsilon}^{2}|s_{i,u}|^{2}}\!\!\right)\!\!\right]
 \nonumber\\
 &=&\!\!\!\!\! \Pr (\widetilde{D}>0),
    \label{eq:PEP_Ge_est}
\end{eqnarray}
\end{small}
\hrulefill
\end{figure*}
\setcounter{equation}{28}
\begin{figure*}
    \begin{equation}
    P_{I}(\bm{s} \!\rightarrow \!\hat{\bm{s}}|\widetilde{\bm{H}}_{u},I_{u}^{\prime})
    \!= \!Q\left(\sqrt{\sum_{u=1}^{U}\left( \sum\limits_{i=1}^{I_{u}^{\prime}} \frac{ |\tilde{h}_{l_{i,u}}\Delta_{i,u}|^{2}}{2(\sigma^{2}+\sigma_{J}^{2}+\sigma_{\epsilon}^{2}|s_{i,u}|^{2})}
    +\sum\limits_{i=1+I_{u}^{\prime}}^{I}\frac{|\tilde{h}_{l_{i,u}}\Delta_{i,u}|^{2}}{2(\sigma^{2}+\sigma_{\epsilon}^{2}|s_{i,u}|^{2})}\right)}\right).
   \label{eq:PEP_est}
\end{equation}
\hrulefill
\end{figure*}
\setcounter{equation}{29}
\begin{figure*}
    \begin{eqnarray}
    P_{I,U}^{\text{im}}(\bm{s} \!\rightarrow \!\hat{\bm{s}}|I_{u}^{\prime})\!\!\!\!\!&=&\!\!\!\!\!\frac{\exp\left[\sum\limits_{i=1}^{I}\sum\limits_{u=1}^{U}
    \frac{\frac{\xi\tilde{ \rho}_{i,u} {h}^{2}_{l_{i,u},LoS}}{4(1+\xi)}}{1-\frac{\tilde{ \rho}_{i,u} }{4} {h}^{2}_{l_{i,u},LoS}\left(\frac{\sigma_{i,u}^{2}}{1+\xi}-\sigma_{\epsilon}^{2}\right)}\right]}
    {12\prod\limits_{i=1}^{I}\prod\limits_{u=1}^{U}\left[1\!\!-\!\!\frac{\tilde{ \rho}_{i,u} }{4} {h}^{2}_{l_{i,u},LoS}\left(\frac{\sigma_{i,u}^{2}}{1+\xi}-\sigma_{\epsilon}^{2}\right)\right]}
+
   \frac{\exp\left[\sum\limits_{i=1}^{I}\sum\limits_{u=1}^{U}
    \frac{\frac{\xi\tilde{ \rho}_{i,u} {h}^{2}_{l_{i,u},LoS}}{3(1+\xi)}}{1-\frac{\tilde{ \rho}_{i,u} }{3} {h}^{2}_{l_{i,u},LoS}\left(\frac{\sigma_{i,u}^{2}}{1+\xi}-\sigma_{\epsilon}^{2}\right)}\right]}
    {4\prod\limits_{i=1}^{I}\prod\limits_{u=1}^{U}\left[1\!\!-\!\!\frac{\tilde{ \rho}_{i,u} }{3} {h}^{2}_{l_{i,u},LoS}\left(\frac{\sigma_{i,u}^{2}}{1+\xi}-\sigma_{\epsilon}^{2}\right)\right]},
    \label{eq:PEP_U_est}
\end{eqnarray}
\hrulefill
\end{figure*}

\subsection{ABER Analysis Under Imperfect CSI }~\label{subsec:PEP}
In practical scenarios, it is difficult for channel estimators at the receiver to estimate channels perfectly, thus resulting in channel estimation errors. The estimated channel gain, denoted by $\tilde{h}_{l_{i,u}}$, corresponding to $h_{l_{i,u}}$ is given as follows~\cite{channel_est}:
\begin{equation}
\setcounter{equation}{23}
   \tilde{h}_{l_{i,u}}=h_{l_{i,u}}-\epsilon_{i,u},
\end{equation}
where $\epsilon_{i,u}$ is the channel estimation error with zero mean and variance $\sigma_{\epsilon}^{2}$. It is noticed that $h_{l_{i,u}}$ and $\epsilon_{i,u}$ are independent with each other and the variance of $\epsilon_{i,u}$ refers to the accuracy of channel estimation.

Replacing $h_{l_{i,u}}$ in Eq.~\eqref{eq:yi} by $\tilde{h}_{l_{i,u}}$, we can rewrite the received signal as follows:
\begin{eqnarray}
   y_{i,u}=\underbrace{\tilde{h}_{l_{i,u}}s_{i,u}}_{\text{useful signal}}+\overbrace{\underbrace{\epsilon_{i,u}s_{i,u}}_{\text{estimation error}}+\kappa_{i,u} w_{i,u}+n_{i,u}}^{\text{overall interference}},
   \label{eq:yi_est}
 \end{eqnarray}
where the variance of estimation error is equal to $\sigma_{\epsilon}^{2}|s_{i,u}|^{2}$.

Then, under imperfect CSI, the estimation of $\bm{s}$ by the ML decoder is expressed as follows:
\begin{eqnarray}
    \hat{\bm{s}}
    =\arg \min_{\bm{s}\in \mathcal {S}} &&\hspace{-0.5cm}\left\{\sum_{u=1}^{U}\left(\sum\limits_{i=1}^{I_{u}^{\prime}}\frac{|y_{i,u}-\tilde{h}_{l_{i,u}}s_{i,u}|^{2}}
    {\sigma_{J}^{2}+\sigma^{2}+\sigma_{\epsilon}^{2}|s_{i,u}|^{2}}\right.\right.
\nonumber\\
 &&\left.\left.\hspace{-0.5cm}+\!\!\!\!\sum\limits_{i=1+I_{u}^{\prime}}^{I}\!\!\frac{|y_{i,u}\!-\!\tilde{h}_{l_{i,u}}s_{i,u}|^{2}}{\sigma^{2}+\sigma_{\epsilon}^{2}|s_{i,u}|^{2}}\right)\right\}.
  \label{eq:ml_est}
\end{eqnarray}
In the similar way as deriving Eq.~\eqref{eq:PEP_Ge} and based on Eqs.~\eqref{eq:yi_est} and \eqref{eq:ml_est}, the conditional PEP, denoted by $P_{I}(\bm{s}\!\rightarrow \!\hat{\bm{s}}|\widetilde{\bm{H}}_{u},I_{u}^{\prime})$, under imperfect CSI is derived as shown in Eq.~\eqref{eq:PEP_Ge_est}, where $\widetilde{\bm{H}}_{u}$ is the diagonal channel matrix with the entries $\tilde{h}_{l_{i,u}}$ and $\widetilde{D}$ is the difference between the two sides of unequal sign. 
Thus, the decision variable $\widetilde{D}$ has the mean and variance, denoted by $\widetilde{D}_{\mu}$ and $\widetilde{D}_{var}$, as
\setcounter{equation}{26}
\begin{equation}
    \widetilde{D}_{\mu}\!=\!-\!\!\sum_{u=1}^{U}\!\left(\!\sum\limits_{i=1}^{I_{u}^{\prime}} \frac{ |\tilde{h}_{l_{i,u}}\Delta_{i,u}|^{2}}{\sigma^{2}\!+\!\sigma_{J}^{2}\!+\!\sigma_{\epsilon}^{2}|s_{i,u}|^{2}}
    \!+\!\sum\limits_{i=1+I_{u}^{\prime}}^{I}\!\frac{|\tilde{h}_{l_{i,u}}\Delta_{i,u}|^{2}}{\sigma^{2}\!+\!\sigma_{\epsilon}^{2}|s_{i,u}|^{2}}\!\right)
\end{equation}
and
\begin{equation}
    \widetilde{D}_{var}\!\!=\!2\!\!\sum_{u=1}^{U}\!\!\left(\!\sum\limits_{i=1}^{I_{u}^{\prime}} \frac{ |\tilde{h}_{l_{i,u}}\Delta_{i,u}|^{2}}{\sigma^{2}\!+\!\sigma_{J}^{2}\!+\!\sigma_{\epsilon}^{2}\!|s_{i,u}\!|^{2}}
   \! \! +\!\sum\limits_{i=1+I_{u}^{\prime}}^{I}\frac{|\tilde{h}_{l_{i,u}}\Delta_{i,u}|^{2}}{\sigma^{2}\!+\!\sigma_{\epsilon}^{2}|s_{i,u}|^{2}}\!\!\right),
\end{equation}
respectively~\cite{Fun_wc}. Then, we can obtain the expression of $P_{I}(\bm{s}\!\rightarrow \!\hat{\bm{s}}|\widetilde{\bm{H}}_{u},I_{u}^{\prime})$ as shown in Eq.~\eqref{eq:PEP_est}.
It is clear that Eq.~\eqref{eq:PEP_est} gets back to Eq.~\eqref{eq:PEP_2} if $\sigma_{\epsilon}^{2}=0$, namely perfect CSI. Under imperfect CSI, the mean of $\tilde{h}_{l_{i,u}}$ is $\sqrt{\frac{\xi}{1+\xi}}h_{l_{i,u},LoS}$ while the variance is $\left(\frac{\sigma_{i,u}^{2}}{\xi+1}-\sigma_{\epsilon}^{2}\right)$. Then, following the similar steps in Appendix~\ref{app:P1}, the conditional PEP under imperfect CSI for arbitrary $U^{\prime}$ and $I_{u}^{\prime}$ is obtained as shown in Proposition 2.

{\textit{Proposition 2}:} The exact PEP, denoted by $P_{I,U}^{\text{im}}(\bm{s} \!\rightarrow \!\hat{\bm{s}}|I_{u}^{\prime})$, with $I_{u}^{\prime}$ OAM-modes jammed at each hop under imperfect CSI is given by Eq.~\eqref{eq:PEP_U_est},
where
\setcounter{equation}{30}
\begin{equation}
   \tilde{ \rho}_{i,u}=\left\{
    \begin{aligned}
       &-\frac{\Delta_{i,u}^{2}}{\sigma^{2}+\sigma_{J}^{2}+\sigma_{\epsilon}^{2}|s_{i,u}|^{2}}  \ \ \ \text{if}\ i\in [1,I_{u}^{\prime}];\\
       &-\frac{\Delta_{i,u}^{2}}{\sigma^{2}+\sigma_{\epsilon}^{2}|s_{i,u}|^{2}} \ \ \ \ \ \ \ \ \ \ \text{if}\ i\in [1+I_{u}^{\prime},I].
    \end{aligned}
    \right.
\end{equation}
Replacing $P_{I,U}(\bm{s} \!\rightarrow \!\hat{\bm{s}}|I_{u}^{\prime})$ in Eq.~\eqref{eq:APEP} by $P_{I,U}^{\text{im}}(\bm{s} \!\rightarrow \!\hat{\bm{s}}|I_{u}^{\prime})$, we can obtain the ABER of our proposed IM-MH scheme under imperfect CSI.

The expressions of conditional PEP and jamming probability are the same for our proposed IM-MH scheme and the previously proposed MH scheme when $I=1$. However, our proposed IM-MH scheme has lower ABER as compared with the MH scheme because $N_{e}(\bm{s}, \hat{\bm{s}})/\eta_{0}$ is used in the IM-MH scheme instead of $N_{e}(\bm{s}, \hat{\bm{s}})/\eta_{s}$ in the MH scheme. Therefore, our proposed IM-MH scheme achieves more robust anti-jamming results than the MH scheme.

To significantly increase the SE in both signal information and index information for our proposed IM-MH scheme, activating multiple OAM-modes is required. However, such a requirement leads to the high probability of signals being jammed at each hop when the jammer sends jamming signals with OAM beams, thus causing inefficient anti-jamming results. In order to deal with the above-mentioned problem, a novel IM-DSMH scheme is proposed to effectively protect transmit signals from hostile jamming.

\section{The IM-DSMH Scheme}\label{sec:DMMH}

\begin{figure}
\centering
  \includegraphics[width=0.4\textwidth]{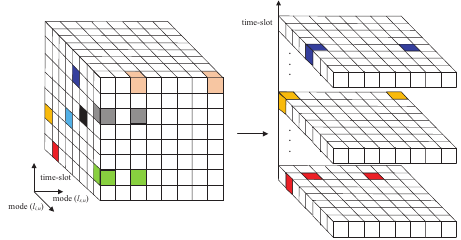}\\
  \caption{An example of IM-DSMH pattern. The time-mode($l_{i,u}$)-mode($l_{s,u}$) resource cubes with specified color denote the activated OAM-modes at each hop. The activated OAM-modes in selectors A and B are rapidly changes according to the preset IM-DSMH pattern. }
  \label{fig:IM_DSMH}
\end{figure}

As shown in Fig.~\ref{fig:sys}, index selector B at the transmitter and additional DFT operator at the receiver are added into the OAM-based index modulation wireless communications system when DSMH works. Thereby, at the transmitter two index selectors work for the IM-DSMH scheme. The index information determines the activated OAM-modes in both index selectors A and B. Based on the index information bits, one OAM-mode is activated as the second hop to transmit the hopping signals in the proposed IM-MH scheme.

To illustrate the hopping of the IM-DSMH scheme, an example of the IM-DSMH pattern is depicted in Fig.~\ref{fig:IM_DSMH}, where the time resource is divided into multiple time slots. The activated OAM-mode in selector B is denoted by $l_{s,u}$ $(|l_{s,u}| \leq N/2)$. We integrate OAM-mode $l_{i,u}$, $l_{s,u}$ and time slot into a three-dimension resource cube. Each hop corresponds to a cube and is identified by the specified color.

Therefore, the output signal, denoted by $s_{l_{s,u}}$, after index selector B for the $u$-th hop is given as follows:
\begin{equation}
    s_{l_{s,u}}=\sum_{i=1}^{I}s_{i,u}e^{j\frac{2 \pi }{N} l_{i,u} l_{s,u}}.
\end{equation}
Next, 
the emitted modulated signal, denoted by $x_{n,l_{s,u}}$, for the $n$-th transmit element is derived as follows:
 \begin{eqnarray}
    x_{n,l_{s,u}}&=& s_{l_{s,u}} e^{j\frac{2 \pi }{N} n l_{s,u}}
    \nonumber\\
    &=&\sum_{i=1}^{I}{s}_{i,u} e^{j\frac{2\pi n}{N}l_{s,u} (l_{i,u}+n)}.
\end{eqnarray}
It is clear that the IM-DSIM scheme still keeps the orthogonality of OAM beams. 

Then, the received signal, denoted by $\tilde{y}_{m,u}$, for the $m$-th element at the $u$-th hop is given as follows:
\begin{equation}
    \tilde{y}_{m,u}=\sum_{n=0}^{N-1}h_{mn}x_{n,l_{s,u}}+ J_{m} + n_{m}.
\end{equation}

We multiply the received signals with $\exp\{-j\frac{2\pi m l_{i,u}}{N} \}$ for the first de-hopping and then average the sum of signals for all receive elements. Thus, we can obtain the de-hopping signal, denoted by ${y}_{l_{i,u}}$, corresponding to OAM-mode $l_{i,u}$ as follows:
\begin{eqnarray}
    {y}_{l_{i,u}}\!\!\!\!\!\!&=&\!\!\!\!\frac{1}{N}\sum_{m=0}^{N-1}\tilde{y}_{m,u}e^{-j\frac{2\pi m l_{i,u}}{N}}
    \nonumber\\
  \!\! \!\!\!\! &=&\!\!\!\! h_{l_{i,u}} s_{l_{i,u}}e^{j\frac{2 \pi }{N} l_{i,u} l_{s,u}}\!\!\!+\!\!\frac{1}{N}\sum_{m=0}^{N-1}(J_{m}\! +\! n_{m})e^{-j\frac{2\pi m l_{i,u}}{N}}.
  \nonumber\\
\end{eqnarray}
To further de-hop the obtained signal, the second DFT algorithm is performed. Thus, we have the de-hopping signal, denoted by $y_{s,u,i}$, for the activated OAM-modes $l_{i,u}$ and $l_{s,u}$ as follows:
\begin{equation}
     y_{s,u,i}=\frac{1}{N}\sum_{i=1}^{I} y_{l_{i,u}}e^{-j\frac{2\pi }{N} l_{s,u} l_{i,u}}.
\end{equation}

Since attackers do not know the legitimate user's hopping pattern in IM-DSMH communications, jamming signals with different OAM-modes can be eliminated completely after multiplying the received signal with $\exp\{-j\frac{2\pi m}{N} l_{i,u}\}$. The residual jamming signal with OAM-mode $l_{i,u}$ has been decomposed from the vorticose beams to plane beams. Then, after multiplying the received signal with $\exp\{-j\frac{2\pi }{N} l_{s,u} l_{i,u}\}$ and performing the DFT algorithm, the jamming signals can be further eliminated. It is noticed that the legitimate user's signals cannot be jammed when the activated OAM-modes in index selector B do not contain OAM-mode zero. Only when jamming signals carry the same OAM-modes $l_{s,u}$ and the user activates the OAM-mode zero in index selector B, the jamming signals can interfere with the desired signals. Therefore, we propose that all OAM-modes except OAM-mode zero can be activated by index selector B with equal probability to mitigate the interference in wireless communications, thus making it impossible for the legitimate user's signals to be jammed.

Hence, we have the signal $ y_{s,u,i}$, corresponding to the OAM-modes $l_{i,u}$ in index selector A and $l_{s,u}$ in index selector B as follows:
\begin{eqnarray}
     y_{s,u,i}=h_{l_{i,u}}s_{i,u}+n_{l_{s,u},i},
    \label{eq:y_ls}
\end{eqnarray}
where $n_{l_{s},i}$ is the corresponding received noise.

Index selector A activates $I$ out of $N-1$ OAM-modes as the first hop while index selector B activates one out of $N$ OAM-modes as the second serial hop. The activated OAM-modes by index selectors A and B are combined into two-dimensional OAM-OAM blocks. Each OAM-OAM combination corresponds to the specific index information. Therefore, the number of available OAM-OAM combinations is $K_{2}=2^{\left\lfloor \log_{2}\left[\binom{N-1}{I}N\right]\right\rfloor}$. Thus, the corresponding transmission bits, denoted by $\eta_{1}$, of the IM-DSMH scheme is expressed as follows:
\begin{eqnarray}
     \eta_{1}&=&I\log_{2}M+U\log_{2}K_{2}.\label{eq:eta_dsmh}
\end{eqnarray}
Comparing the transmission bits of our proposed IM-MH and IM-DSMH schemes, we can obtain the increase of transmission bits, denoted by $\Delta_{\eta}$, for the proposed IM-DSMH scheme as follows:
\begin{eqnarray}
    \Delta_{\eta}&=&\eta_{1}-\eta_{0}
    \nonumber\\
    &\overset{(a)}{\approx} & U\log_{2}\left(N-I\right),
    \label{eq:eta_di}
\end{eqnarray}
where $(a)$ ignores the floor operation. It is clear that $\Delta_{\eta}$ increases monotonically with the increase of $N$ and $U$, respectively.

The derivation of ABER for our proposed IM-DSMH scheme is similar to that for the IM-MH scheme. By considering all possible cases, the ABER, denoted by $\tilde{P}_{I,U}$, with $I$ activated OAM-modes and $U$ hops for our proposed IM-DSMH scheme is derived with the union-bound method~\cite{Digital_C} as follows:
\begin{equation}
   \tilde{P}_{I,U} \leq \sum\limits_{\bm{s}}^{2^{\eta_{s}}}
   \sum\limits_{\hat{\bm{s}}}^{2^{\eta_{s}}}\frac{\tilde{P}_{I,U}(\bm{s} \rightarrow \hat{\bm{s}})N_{e}(\bm{s}, \hat{\bm{s}})}{\eta_{1}2^{\eta_{s}}},
\end{equation}
where $\tilde{P}_{I,U}(\bm{s} \rightarrow \hat{\bm{s}})$ is expressed as
    \begin{eqnarray}
  \!\!\!\! \tilde{P}_{I,U} (\bm{s} \!\rightarrow \!\hat{\bm{s}})\!\!\!\!&=& \!\!\!\!\frac{\exp\left(-\sum\limits_{u=1}^{U}\sum\limits_{i=1}^{I}\frac{\xi\Delta_{i,u}^{2}{h}^{2}_{l_{i,u},LoS}}
   {4(1+\xi)\sigma^{2}+\sigma_{i,u}^{2}\Delta_{i,u}^{2}{h}^{2}_{l_{i,u},LoS}}\right)}
   {12\prod\limits_{u=1}^{U}\prod\limits_{i=1}^{I}\left[1+\frac{\sigma_{i,u}^{2}\Delta_{i,u}^{2}{h}^{2}_{l_{i,u},LoS}}{4(1+\xi)\sigma^{2}}\right]}
   \nonumber\\
 \!\!\!\!  &+& \!\!\!\!\frac{\exp\left(-\sum\limits_{u=1}^{U}\sum\limits_{i=1}^{I}\frac{\xi\Delta_{i,u}^{2}{h}^{2}_{l_{i,u},LoS}}
   {3(1+\xi)\sigma^{2}+\sigma_{i,u}^{2}\Delta_{i,u}^{2}{h}^{2}_{l_{i,u},LoS}}\right)}
   {4\prod\limits_{u=1}^{U}\sum\limits_{i=1}^{I}\left[1+\frac{\sigma_{i,u}^{2}\Delta_{i,u}^{2}{h}^{2}_{l_{i,u},LoS}}{3(1+\xi)\sigma^{2}}\right]}
   \nonumber\\
\end{eqnarray}
under perfect CSI and
\begin{eqnarray}
    \tilde{P}_{I,U}(\bm{s} \!\rightarrow \!\hat{\bm{s}})\!\!\!\!&=&\!\!\!\!\frac{\exp\left[-\sum\limits_{u=1}^{U}\sum\limits_{i=1}^{I}
    \frac{\frac{\xi\Delta_{i,u}^{2} {h}^{2}_{l_{i,u},LoS}}{4(1+\xi)(\sigma^{2}+\sigma_{\epsilon}^{2}|s_{i,u}|^{2})}}{1+\frac{ \Delta^{2}_{i,u} {h}^{2}_{l_{i,u},LoS}}{4(\sigma^{2}+\sigma_{\epsilon}^{2}|s_{i,u}|^{2})} \left(\frac{\sigma_{i,u}^{2}}{1+\xi}-\sigma_{\epsilon}^{2}\right)}\right]}
    {12\prod\limits_{u=1}^{U}\prod\limits_{i=1}^{I}\left[1\!\!+\!\!\frac{\Delta^{2}_{i,u} {h}^{2}_{l_{i,u},LoS} }{4(\sigma^{2}+\sigma_{\epsilon}^{2}|s_{i,u}|^{2})} \left(\frac{\sigma_{i,u}^{2}}{1+\xi}-\sigma_{\epsilon}^{2}\right)\right]}
    \nonumber\\
   &+&\!\!\!\!\!\!
 \frac{\exp\left[-\sum\limits_{u=1}^{U}\sum\limits_{i=1}^{I}
    \frac{\frac{I\xi\Delta_{i,u}^{2} {h}^{2}_{l_{i,u},LoS}}{3(1+\xi)(\sigma^{2}+\sigma_{\epsilon}^{2}|s_{i,u}|^{2})}}{1+\frac{ \Delta^{2}_{i,u} {h}^{2}_{l_{i,u},LoS}}{3(\sigma^{2}+\sigma_{\epsilon}^{2}|s_{i,u}|^{2})} \left(\frac{\sigma_{i,u}^{2}}{1+\xi}-\sigma_{\epsilon}^{2}\right)}\right]}
    {4\prod\limits_{u=1}^{U}\prod\limits_{i=1}^{I}\left[1\!\!+\!\!\frac{\Delta^{2}_{i,u} {h}^{2}_{l_{i,u},LoS} }{3(\sigma^{2}+\sigma_{\epsilon}^{2}|s_{i,u}|^{2})} \left(\frac{\sigma_{i,u}^{2}}{1+\xi}-\sigma_{\epsilon}^{2}\right)\right]}
    \nonumber\\
\end{eqnarray}
under imperfect CSI, respectively. It is noticed that our proposed schemes can use various modulation schemes, such as phase-shift keying (PSK), quadrature amplitude modulation (QAM), and frequency-shift keying (FSK).

The proposed IM-MH scheme not only simultaneously activates several OAM-modes for transmission at each hop according to the input index information, but also utilizes the diversity of fast hopping to decrease the probability of signals being jammed. The proposed IM-DSMH scheme with slight computational complexity cost uses the double-serial hop to increase the index information and effectively avoid legitimate signals being jammed. Thus, our proposed IM-MH and IM-DSMH schemes within a narrow frequency band can be used for achieving high SE and low ABER for various interfering waveforms such as single-tone interference, wideband interference, and partial-band interference. Therefore, our proposed IM-MH scheme can be applied into the scenarios, such as indoor wireless communication, radar, wireless local area networks.

\vspace{-5pt}

\section{PERFORMANCE EVALUATION}~\label{sec:simu}
In this section, we show several numerical and simulation results to evaluate the performance of our proposed IM-MH and IM-DSMH schemes.
The numerical results are organized as follows. Section~\ref{sec:ABER_MH} shows the ABERs of our proposed IM-MH scheme. Section~\ref{sec:ABER_DSMH} depicts the ABERs of our proposed IM-DSMH scheme and compares the ABERs among IM-MH, IM-DSMH, FH, and MH schemes. Section~\ref{sec:SE_comparison} compares the SEs among IM-MH, IM-DSMH, FH, and MH schemes. Throughout our evaluations, the jamming-to-noise ratio is set to 2 dB and binary PSK (BPSK) modulation is employed.

\subsection{ABER of IM-MH Scheme}\label{sec:ABER_MH}

\begin{figure}
\centering
  \includegraphics[width=0.42\textwidth]{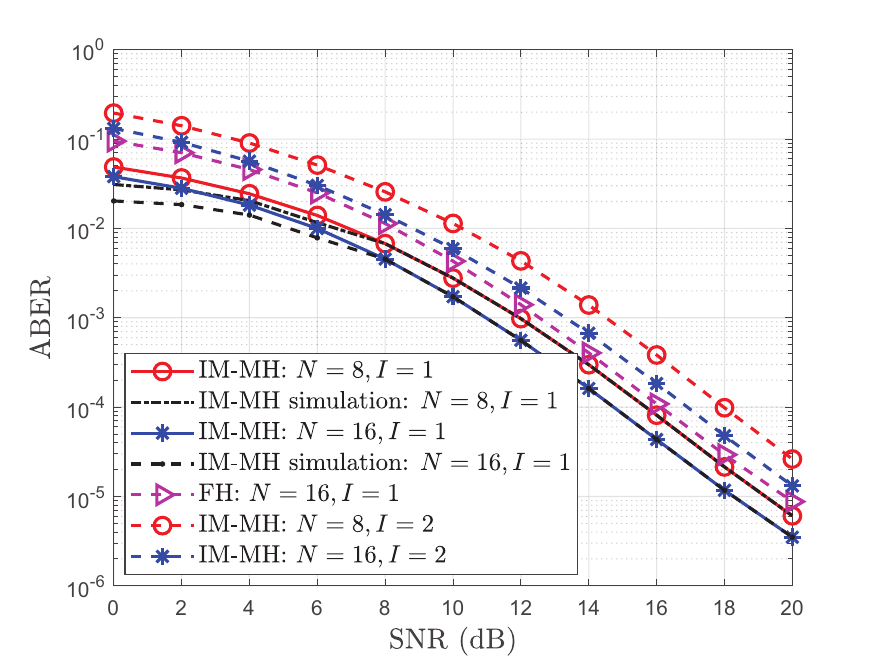}\\
  \caption{The impact of $N$ and $I$ on the ABERs of IM-MH and FH schemes under perfect CSI.}
  \label{fig:MH_BER_perfect_NI}
  \vspace{-5pt}
\end{figure}
To evaluate the impact of $N$ and $I$ on the ABERs of the IM-MH and FH schemes under perfect CSI, we present the ABER versus signal-to-noise ratio (SNR) in Fig.~\ref{fig:MH_BER_perfect_NI}, where we set $\xi=10$ and $U=1$. The ABER corresponding to $N=8$ and $I=2$ is the highest among the four cases because the jamming probability is the highest among the four cases as proved in Eq.~\eqref{eq:jamming_p}. The other reason is that the case for $I=2$ has a denser joint constellation size than that for $I=1$, which makes it more possible for the ML decoder to make error decisions. Hence, the ABER of our proposed IM-MH scheme increases as the number of available OAM-modes decreases and the number of activated OAM-modes increases. Also, we can find that for a fixed $I$, the difference between the cases $N=16$ and $N=8$ is larger in a relatively high SNR region than that in the low SNR region. This is because interference plays a dominant role in the low SNR region while legitimate signals play a dominant role in the high SNR region. In addition, comparing the ABERs of our proposed IM-MH and the conventional FH schemes, we can find that our proposed IM-MH scheme achieves lower ABER than the FH scheme. This is because our proposed IM-MH scheme can simultaneously transmit index bits and signal bits. The index bits related to activated OAM-modes can be easily obtained at the receiver due to the pre-shared secret keys between the transmitter and receiver. Therefore, when given the transmission bits, the IM-MH scheme has a lower ABER than the FH scheme.

\begin{figure}
\centering
  \includegraphics[width=0.42\textwidth]{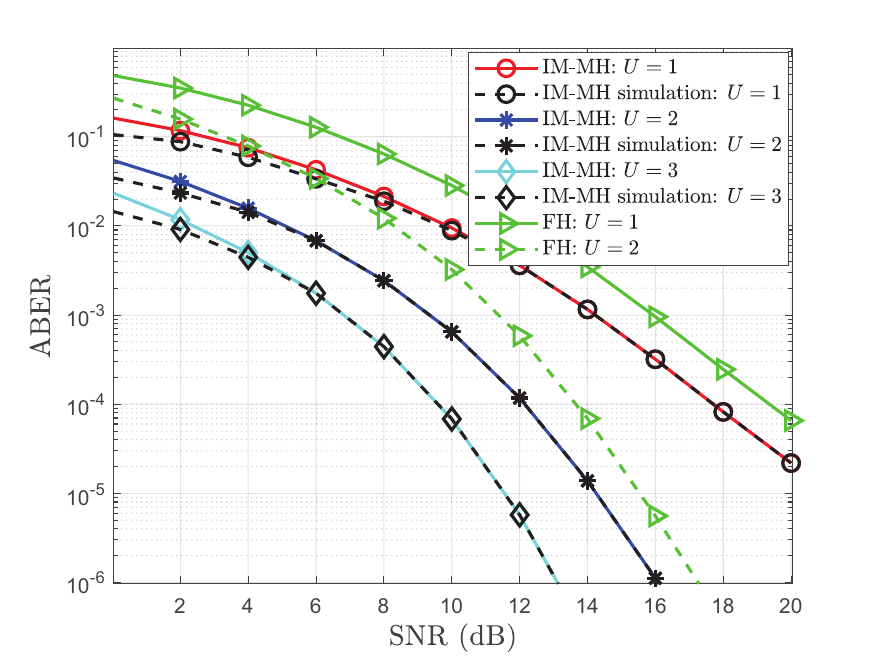}\\
  \caption{The impact of $U$ on the ABERs of IM-MH and FH schemes with $I=2$ and $N=8$ under perfect CSI.}
  \label{fig:MH_BER_U_NI_perfect}
  \vspace{-5pt}
\end{figure}

\begin{figure}
\centering
  \includegraphics[width=0.42\textwidth]{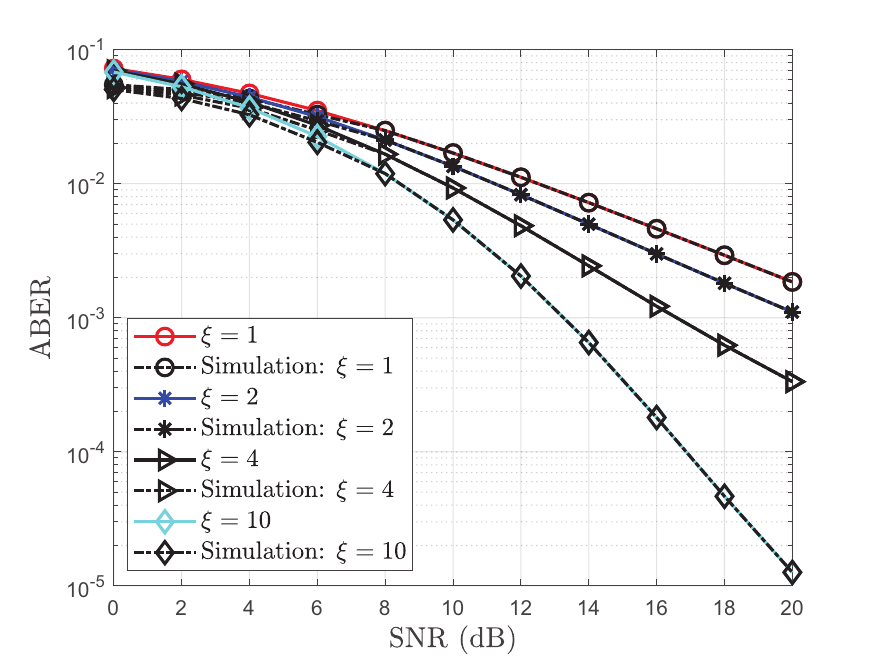}\\
  \caption{The ABERs of our proposed IM-MH scheme versus SNR for different values of Rician factor $\xi$ under perfect CSI. }
  \label{fig:MH_BER_U1_NI_perfect_xi}
  \vspace{-5pt}
\end{figure}

Figure~\ref{fig:MH_BER_U_NI_perfect} compares the ABERs of our proposed IM-MFH and the conventional FH schemes versus different values of $U$, where we set $N=8$, $I=2$, and $\xi=10$, respectively. As expected, the ABER monotonically decreases as the number of hops increases, thus resulting in decreasing the jamming probability as verified in Eq.~\eqref{eq:jp}. This implies that it is difficult for attackers to track trajectories of activated OAM-modes. Also, modulated signals are transmitted within $U$ hops duration, thus bringing a diversity gain of wireless communications to improve the anti-jamming performance.
This result verifies that by increasing the number of hops, our proposed IM-MH scheme can achieve robust anti-jamming in wireless communications. Observing Fig.~\ref{fig:MH_BER_U_NI_perfect}, we can find that the ABER of our proposed IM-MH scheme is lower than that of the conventional FH scheme because our proposed IM-MH scheme increases the index bits for the fixed $U$ and $I$.

Figure~\ref{fig:MH_BER_U1_NI_perfect_xi} shows the ABERs of our proposed IM-MH scheme versus channel SNR for different values of $\xi$ under perfect CSI, where we set $N=4$, $I=4$, and $U=1$, respectively. Observing Fig.~\ref{fig:MH_BER_U1_NI_perfect_xi}, we can obtain that the ABER decreases as the value of $\xi$ increases. This is because the LoS component plays a dominant role in the high $\xi$ region whereas NLoS components dominate in the low $\xi$ region. When $\xi$ is high, the impact of multipath fading on ABER is mitigated.

\begin{figure}
\centering
  \includegraphics[width=0.42\textwidth]{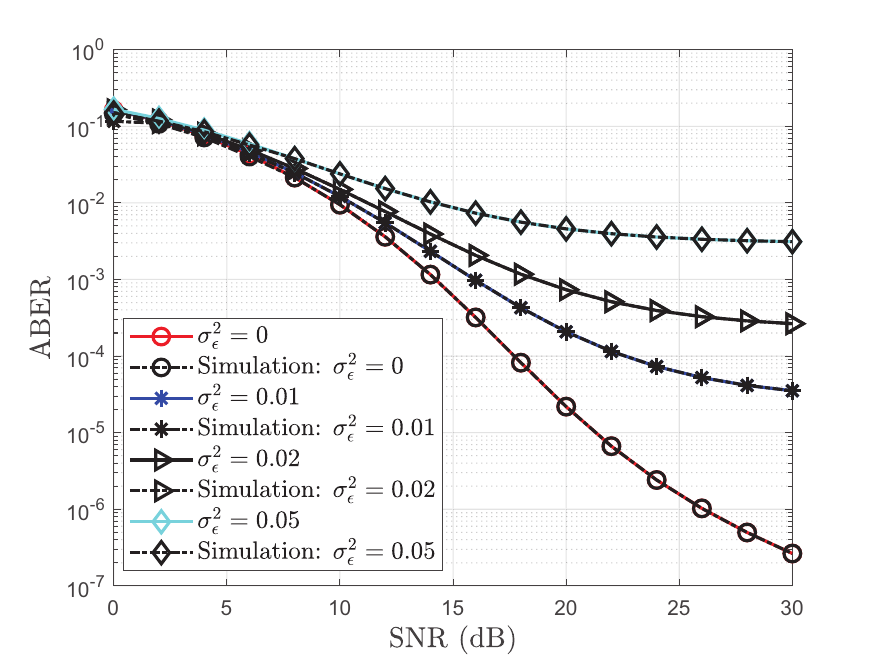}\\
  \caption{The impact of $\sigma_{\epsilon}^{2}$ on the ABERs of our proposed IM-MH scheme.}
  \label{fig:MH_BER_U1_NI_imperfect_coefficient}
  \vspace{-5pt}
\end{figure}

Figure~\ref{fig:MH_BER_U1_NI_imperfect_coefficient} depicts the impact of channel estimation error on the ABERs of our proposed IM-MH scheme versus the received SNR, where we set $N=8$, $I=2$, and $\xi=10$, respectively. As shown in Eq.~\eqref{eq:yi_est}, $\sigma_{\epsilon}^{2}$ refers to the accuracy of channel estimation. $\sigma_{\epsilon}^{2}=0$ means that the CSI is estimated perfectly. As $\sigma_{\epsilon}^{2}$ increases, the received SINR decreases, thus resulting in high ABER. Therefore, as shown in Fig.~\ref{fig:MH_BER_U1_NI_imperfect_coefficient}, the ABER curve for $\sigma_{\epsilon}^{2}=0$ can be considered as the ABER threshold for imperfect CSI scenarios. As the channel SNR increases, the ABER first slowly decreases and then gets close to a fixed value as shown in Fig.~\ref{fig:MH_BER_U1_NI_imperfect_coefficient}. However, there is no fixed value of ABER under perfect CSI in the whole SNR region.

\begin{figure}
\centering
 \includegraphics[width=0.42\textwidth]{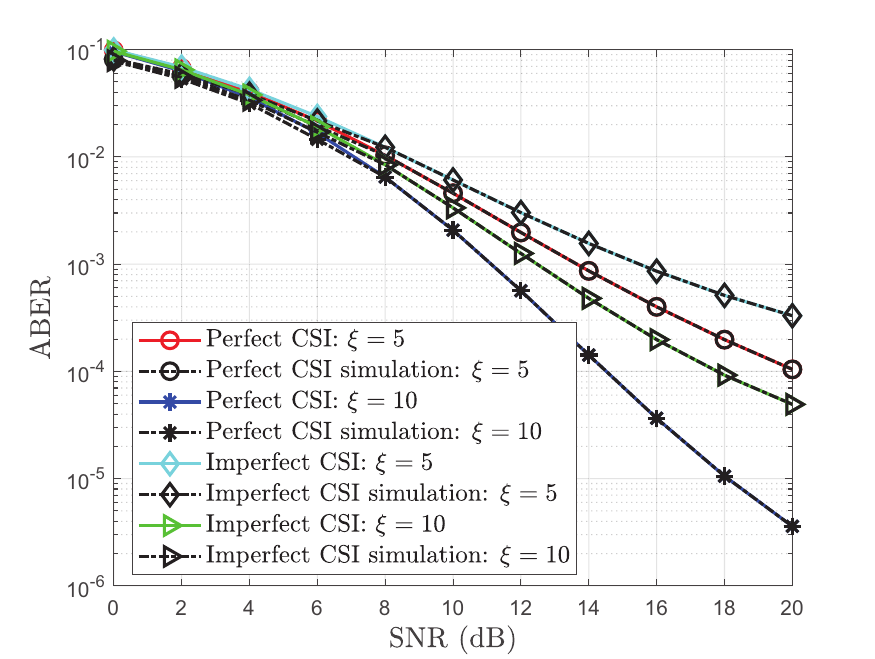}\\
\caption{The ABERs of our proposed IM-DSMH scheme for different values of $\xi$ with perfect and imperfect CSI.}
  \label{fig:DSMH_BER_XI}
  \vspace{-5pt}
\end{figure}

\begin{figure}
\centering
  \includegraphics[width=0.42\textwidth]{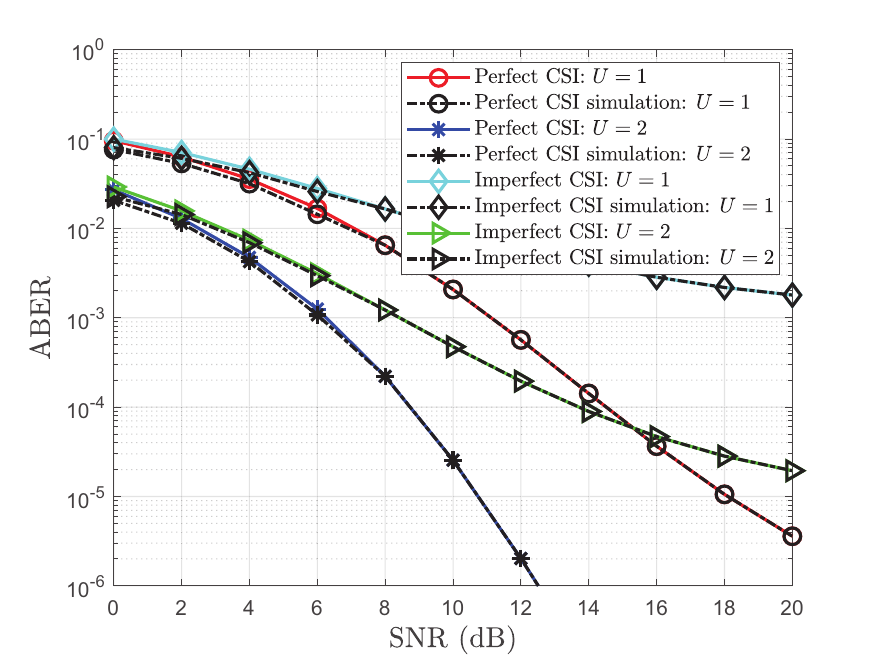}\\
  \caption{The effect of $U$ on the ABERs of our proposed IM-DSMH scheme under perfect and imperfect CSI.}
  \label{fig:DSMH_BER_U}
  \vspace{-5pt}
\end{figure}

\subsection{ABER of IM-DSMH Scheme and ABER Comparisons}\label{sec:ABER_DSMH}
Figures~\ref{fig:DSMH_BER_XI} and \ref{fig:DSMH_BER_U} show the ABERs of our proposed IM-DSMH scheme versus the received SNR for different values of $U$ and $\xi$ under both perfect and imperfect CSI scenarios, where we set $I=2$ and $N=8$, respectively. Observing the ABER curves, we can find that the ABERs under perfect CSI are lower than those under imperfect CSI scenarios. The ABER difference between perfect CSI and imperfect CSI is larger in a relatively high SNR region as compared with that in a low SNR region. Fig.~\ref{fig:DSMH_BER_XI}, where we set $U=1$, shows that the ABERs of our proposed IM-DSMH scheme decreases as $\xi$ increases. Also, observing Fig.~\ref{fig:DSMH_BER_U}, we can see that the ABERs of our proposed IM-DSMH scheme decrease as $U$ increases. This verifies that our proposed IM-DSMH scheme with multiple hops can achieve robust anti-jamming of wireless communications.

\begin{figure}
\centering
  \includegraphics[width=0.42\textwidth]{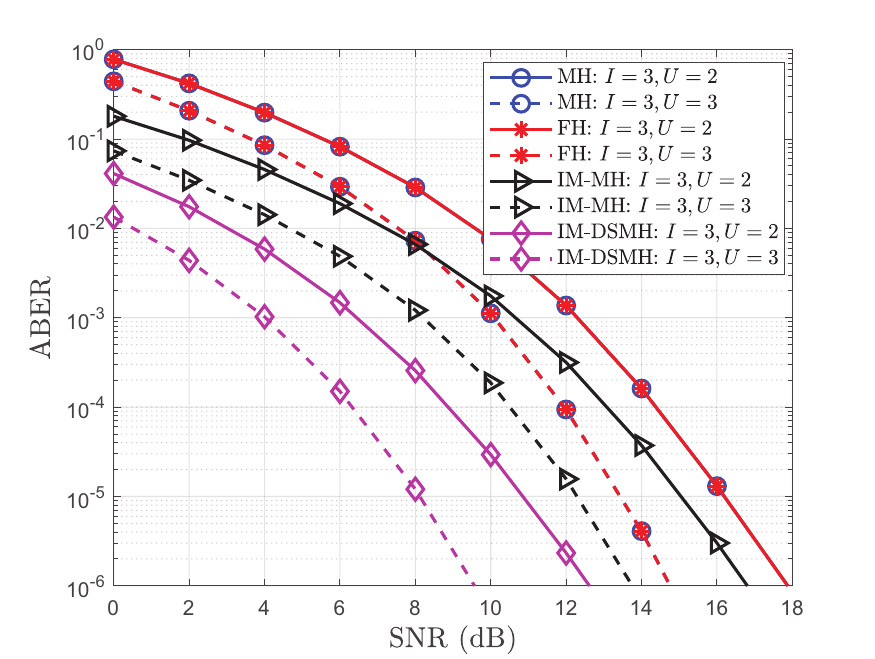}\\
  \caption{ABER comparisons among IM-MH, IM-DSMH, FH, and MH schemes for $N=8$ and $I=3$ under perfect CSI.}
  \label{fig:Comparison_MH_DSMH_MH_perfect}
  \vspace{-5pt}
  \end{figure}
Figure~\ref{fig:Comparison_MH_DSMH_MH_perfect} compares the ABERs of our proposed IM-MH, IM-DSMH, MH, and the conventional FH schemes versus the received SNR for different values of $U$ and $I$ under perfect CSI, where we set $N=8$, $I=3$, and $\xi=10$, respectively. Observing the curves, we can see that our proposed IM-MH and IM-DSMH schemes have lower ABERs than the MH scheme for given $U$ and $I$. Although our proposed IM-MH scheme and the MH scheme have the same jamming probabilities for given $U$ and $I$, our proposed IM-MH scheme has a lower ratio between the number of bit errors and all transmission bits due to the increase of $\eta_{x}$ in comparison with the MH scheme. The MH and FH schemes have the same ABER due to the same jamming probabilities and transmission bits. Also, the IM-DSMH scheme not only has the lowest ratio between the number of bit errors and all transmit bits, but also successfully avoids jamming attacks. Therefore, our proposed IM-DSMH scheme achieves the lowest ABER among the four schemes as shown in Fig.~\ref{fig:Comparison_MH_DSMH_MH_perfect}. In addition, the ABERs of our proposed IM-MH and IM-DSMH schemes are around 25\% and 10\%, respectively, compared with that of the MH scheme.

\begin{figure}
\centering
  \includegraphics[width=0.42\textwidth]{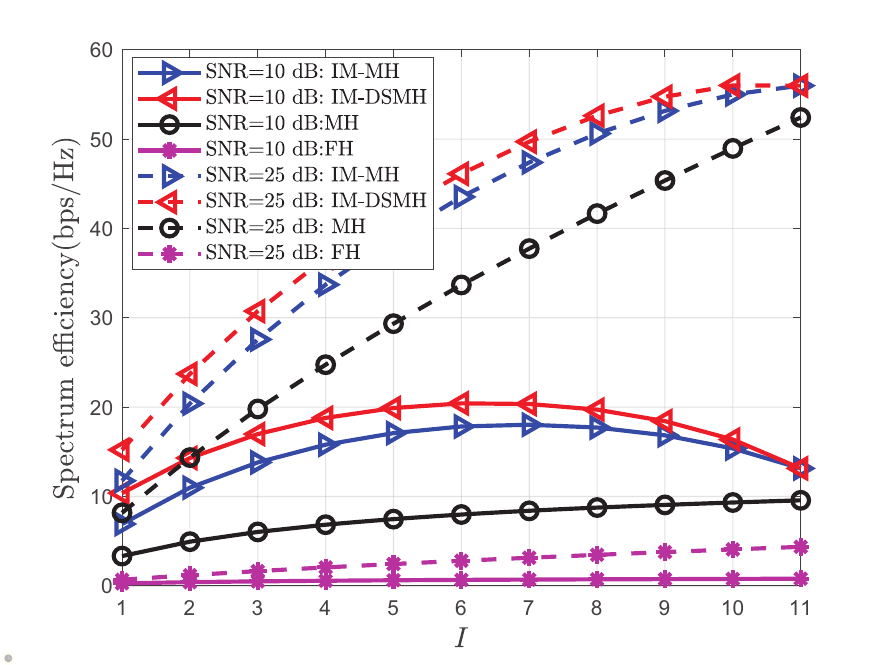}\\
  \caption{SE comparisons among IM-MH, IM-DSMH, FH, and MH schemes versus $I$ with $U=1$.}
  \label{fig:se_comparison_SNR}
  \vspace{-5pt}
  \end{figure}

\subsection{Spectrum Efficiency Comparison}\label{sec:SE_comparison}

Similar to the derivation analysis about SE in~\cite{7317720}, the upper bound of SE, denoted by $C_{1}$, for our proposed IM-MH scheme can be obtained as follows:
\begin{eqnarray}
  C_{1}=\sum_{U^{\prime}=0}^{U}P(U^{\prime}|U)\sum_{i=1}^{I}\log_{2}\left(1+\sum_{u=1}^{U}\gamma_{i,u}\right)+\log_{2}K_{1},
  \label{eq:c_mh}
\end{eqnarray}
where $\gamma_{i,u}$ is the signal-to-interference-plus-noise ratio (SINR) for the $i$-th activated OAM-mode at the $u$-th hop in the IM-MH communication and a maximal-ratio combiner is used at the receiver. The first term on the right hand of Eq.~\eqref{eq:c_mh} is the signal information and the second term corresponds to the index information.

The upper bound of SE, denoted by $C_{2}$, for our proposed IM-DSMH scheme is derived as follows:
\begin{eqnarray}
  C_{2}=\sum_{i=1}^{I}\log_{2}\left(1+\sum_{u=1}^{U}\chi_{i,u}\right)+\log_{2}K_{2},
  \label{eq:c_ds}
\end{eqnarray}
where $\chi_{i,u}$ is the SNR for the $i$-th activated OAM-mode at the $u$-th hop in the IM-DSMH communication.

 Figure~\ref{fig:se_comparison_SNR} compares the SEs of our proposed IM-MH, IM-DSMH, FH and MH schemes with different values of $I$, where $N=12$, $U=1$, and $\xi=10$. To analyze the impact of SNR on SE, we set the SNR as 10 dB and 25 dB, respectively. The SE of our proposed IM-MH and IM-DSMH scheme is upper bounded as the sum of index information and signal information, whereas the SE of MH scheme is obtained through signal information. As shown in Fig.~\ref{fig:se_comparison_SNR}, the SEs of our proposed IM-MH and IM-DSMH schemes first increase and then decrease as $I$ increases when SINR is low. Thus, in this case there exist optimal solutions $I$ to achieve the highest SE. On the one hand, the index information is a concave function with respect to $I$ for fixed $N$. On the other hand, the signal information increases as $I$ increases. Thereby, the SE first increase. When the signal information increases slower than the index information decreases, the SEs begin to decrease. Hence, when SNR is low, the SE curves of our proposed IM-MH and IM-DSMH schemes are concave with respect to $I$. When SNR is relatively high, the SEs of our proposed IM-MH and IM-DSMH schemes monotonically increase as $I$ increases. The reason is that the increase of signal information is always larger than the decrease of index information. Observing Fig.~\ref{fig:se_comparison_SNR}, we also can find that when $I=N-1$, the SEs of our proposed IM-MH and IM-DSMH scheme are the same, which is due to the same index information. The SE of our proposed IM-DSMH scheme is larger than that of the IM-MH scheme because the IM-DSMH scheme provides additional index information in comparison with the IM-MH scheme. This result can be proved by Eq.~\eqref{eq:eta_di}. In addition, we can see that our proposed IM-MH and IM-DSMH schemes achieve higher SE than the MH scheme in wireless communications. It is clear that the FH scheme has the lowest SE among these schemes. This is because our proposed IM-MH, IM-DSMH, and MH schemes can be used for anti-jamming within the narrow frequency band while the FH scheme requires $N$ narrow frequency bands for hopping.

\begin{figure}
\centering
  \includegraphics[width=0.42\textwidth]{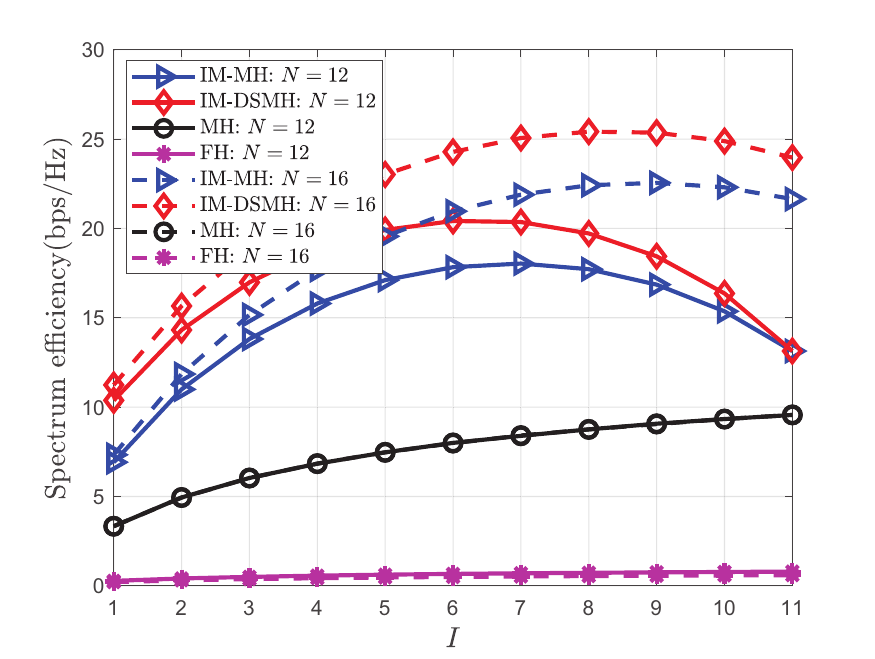}\\
  \caption{The impact of $N$ on SEs among IM-MH, IM-DSMH, FH, and MH schemes. }
  \label{fig:se_comparison_N}
  \end{figure}

  \setcounter{equation}{45}
\begin{figure*}[ht]
    \begin{equation}
    P_{I,U}(\bm{s} \!\rightarrow \!\hat{\bm{s}}|\bm{H}_{u},I_{u}^{\prime})=\frac{1}{12}\exp \!\left(- \!\frac{ \sum\limits_{i=1}^{I_{u}^{\prime}} |h_{l_{i,u}}\Delta_{i,u}|^{2} }{4(\sigma^{2}+\sigma_{J}^{2})}\!-\!\frac{ \sum\limits_{i=1+I_{u}^{\prime}}^{I} |h_{l_{i,u}}\Delta_{i,u}|^{2} }{4\sigma^{2}} \!\right)
    \!\!+\!\!
    \frac{1}{4}\exp \!\left( \!- \!\frac{ \sum\limits_{i=1}^{I_{u}^{\prime}} |h_{l_{i,u}}\Delta_{i,u}|^{2} }{3(\sigma^{2}+\sigma_{J}^{2})}-\frac{ \!\!\sum\limits_{i=1+I_{u}^{\prime}}^{I} |h_{l_{i,u}}\Delta_{i,u}|^{2} }{3\sigma^{2}} \!\right).
    \label{eq:Q_app}
\end{equation}
\hrulefill
\end{figure*}

Figure~\ref{fig:se_comparison_N} compares the SEs of our proposed IM-MH, IM-DSMH, MH, and the conventional FH schemes versus $I$, where $N$ is set as 12 and 16, respectively. As $N$ increases, the SEs of our proposed schemes increase due to the increase of index information as shown in Eqs.~\eqref{eq:eta} and \eqref{eq:eta_dsmh}. It is clear that the SE curves of the MH scheme are overlapped because the SEs are obtained through signal information without considering the jamming attack. Since the index information is always equal to zero for conventional FH and MH schemes, the SEs of our proposed IM-MH and IM-DSMH schemes are larger than those of conventional FH and MH schemes for a fixed $I$.

\section{Conclusions} \label{sec:conc}
In this paper, we proposed the IM-MH scheme to increase the SE and decrease the ABER for anti-jamming in wireless communications within a narrowband. The exact closed-form expressions of ABER were theoretically analyzed under perfect CSI scenarios and imperfect CSI scenarios, respectively. We also proposed the IM-DSMH scheme, which randomly activated an OAM-mode as the second hop to transmit the hopping signals in the IM-MH scheme, to further decrease the ABER. Numerical results have shown that the ABER of our proposed IM-MH scheme within a narrowband decreases as the number of available OAM-modes increases and the number of activated OAM-modes decreases. Our proposed IM-DSMH scheme can achieve robust anti-jamming by performing the second serial hop. In addition, our proposed schemes outperform the MH scheme in terms of SE and ABER within the narrow frequency band.



\begin{appendices}
\section{Proposition 1}\label{app:P1}
We first analyze the conditional PEP for $U=1$ and then for an arbitrary $U$. Thus, Eq.~\eqref{eq:PEP_2} for $U=1$ can be re-expressed by
\setcounter{equation}{44}
\begin{eqnarray}
 \!\!\!\!\!   &&P_{I,U}(\bm{s} \!\rightarrow \!\hat{\bm{s}}|\bm{H}_{u},I_{u}^{\prime})
    \nonumber\\
\!\!\!\!    &&\!=\!Q\!\left(\!\sqrt{\frac{ \sum\limits_{i=1}^{I_{u}^{\prime}} |h_{l_{i,u}}\Delta_{i,u}|^{2}}{2(\sigma^{2}+\sigma_{J}^{2})}
    +\frac{\sum\limits_{i=1+I_{u}^{\prime}}^{I}|h_{l_{i,u}}\Delta_{i,u}|^{2}}{2\sigma^{2}}}\right).
    \label{eq:cpep}
\end{eqnarray}

Based on Eqs.~\eqref{eq:Q_func} and \eqref{eq:cpep}, we can derive $P_{I,U}(\bm{s} \!\rightarrow \!\hat{\bm{s}}|\bm{H}_{u},I_{u}^{\prime})$ as shown in Eq.~\eqref{eq:Q_app}.
Since the channels follow the independent Rician fading, the conditional PEP $P_{I,U}(\bm{s} \!\rightarrow \!\hat{\bm{s}}|I_{u}^{\prime})$ can be derived as follows:
\setcounter{equation}{46}
\begin{eqnarray}
\!\!\!\!\!\!\!\! P_{I,U}(\bm{s} \!\rightarrow \!\hat{\bm{s}}|I_{u}^{\prime})
 \hspace{-0.3cm}&=&\hspace{-0.4cm}\underbrace{\int_{0}^{\infty}\!\!  \int_{0}^{\infty}\!\! \cdots\!\! \int_{0}^{\infty}}_{I-fold}  P_{I}(\bm{s} \!\rightarrow \!\hat{\bm{s}}|\bm{H}_{u},I_{u}^{\prime})
 \nonumber\\
\!\!\!\!\!\!\!\!  &&\hspace{-0.4cm}\times f\!\left(h_{_{l_{1,u}}}\!\right)\!\cdots\! f\left(h_{l_{I,u}}\!\right)d h_{l_{1,u}} \cdots  d h_{l_{I,u}},
  \label{eq:PEP_h}
\end{eqnarray}
where $f\left(h_{l_{i,u}}\right)$ is the PDF of $h_{l_{i,u}}$. Then, $P_{I,U}(\bm{s} \!\rightarrow \!\hat{\bm{s}}|I_{u}^{\prime})$ with $I$-fold integral can be calculated using the moment generating function (MGF)~\cite{Digital_C}.
Thus, Eq.~\eqref{eq:PEP_h} is calculated as follows:
\begin{equation}
    P_{I,U}(\bm{s} \!\rightarrow \!\hat{\bm{s}}|I_{u}^{\prime})=\!\frac{1}{12}\prod_{i=1}^{I}\! \mathcal {M}_{h_{l_{i,u}}^{2}}\!\!\left(\frac{\rho_{i,u}}{4}\!\!\right)\!
 \!+\!\frac{1}{4}\prod_{i=1}^{I} \mathcal {M}_{h_{l_{i,u}}^{2}}\!\left(\!\frac{\rho_{i,u}}{3}\right),
    \label{eq:PEP_MGF}
\end{equation}
where
\begin{eqnarray}
 \mathcal {M}_{h_{l_{i,u}}^{2}}\left(\frac{\rho_{i,u}}{4}\right) = \int_{0}^{\infty} \exp\left(\frac{-\rho_{i,u} h_{l_{i,u}}^{2}}{4}\right) f(h_{l_{i,u}}) d h_{l_{i,u}}
\end{eqnarray}
is the MGF of random variable $h_{l_{i,u}}^{2}$ at $\frac{\rho_{i,u}}{4}$.

The mean and variance of OAM-based wireless channel $h_{l_{i,u}}$ are $\sqrt{\frac{\xi}{1+\xi}}h_{l_{i,u},LoS}$ and $\frac{\sigma_{i,u}^{2}}{1+\xi}$, respectively. Thus, the MGF associated with Rician fading channel model at $\frac{\rho_{i,u}}{4}$ is further derived as follows~\cite{fading_2005}:
\begin{eqnarray}
\mathcal {M}_{h_{l_{i,u}}^{2}}\left(\frac{\rho_{i,u}}{4}\right)=\frac{(1+\xi)\exp\left(\frac{\xi{h}_{l_{i,u},LoS}^{2}\frac{\rho_{i,u}}{4}}{1+\xi-{h}_{l_{i,u},LoS}^{2}\frac{\rho_{i,u}\sigma_{i,u}^{2}}{4}}\right)}{1+\xi-\frac{\rho_{i,u}\sigma_{i,u}^{2}}{4} {h}_{l_{i,u},LoS}^{2}} .
\end{eqnarray}
Based on Eqs.~\eqref{eq:Q_app} and \eqref{eq:PEP_MGF}, the conditional PEP $P_{I,U}(\bm{s} \!\rightarrow \!\hat{\bm{s}}|I_{u}^{\prime})$ for $U=1$ can be obtained. In the similar way as deriving $P_{I,U}(\bm{s} \!\rightarrow \!\hat{\bm{s}}|I_{u}^{\prime})$ for $U=1$, we can obtain the exact closed-form expression of $P_{I,U}(\bm{s} \!\rightarrow \!\hat{\bm{s}}|I_{u}^{\prime})$ for arbitrary $U$ as shown in Eq.~\eqref{eq:PEP_U}.
\end{appendices}
\bibliographystyle{IEEEbib}
\bibliography{References}
\begin{IEEEbiography}[{\includegraphics[width=1in,height=1.25in,clip,keepaspectratio]{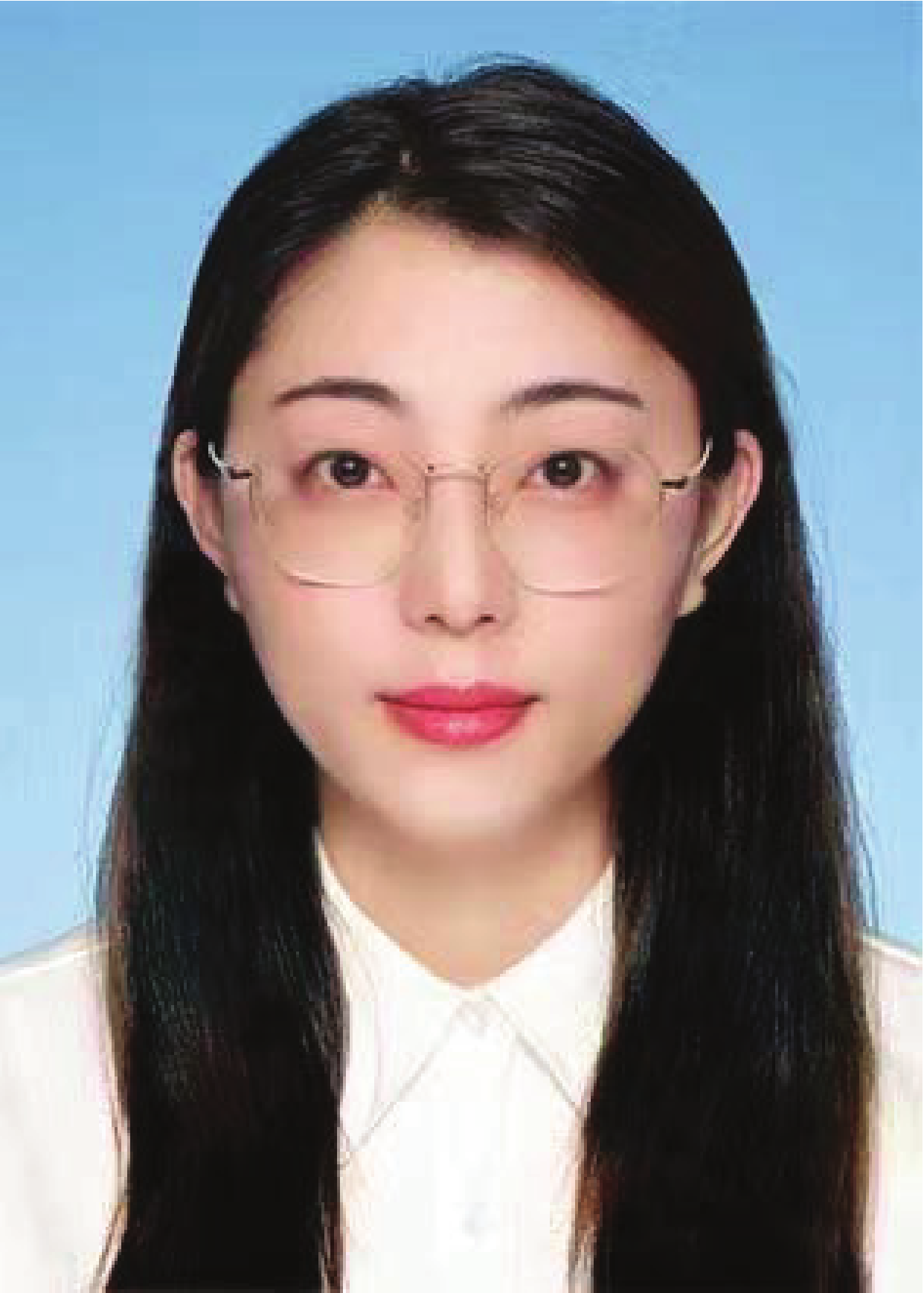}}]{Liping Liang} (SM'18) received the B.S. degree in electronic and information engineering from Jilin University, Changchun, China, in 2015. She received the Ph.D. degree in telecommunication engineering from Xidian University, Xian, China, in 2021. She is currently a lecturer with Xidian University. Her research interests focus on B5G/6G wireless communications with an emphasis on radio vortex wireless communications, integrated sensing and communication, and anti-jamming communications.
\end{IEEEbiography}
\begin{IEEEbiography}[{\includegraphics[width=1in,height=1.5in,clip,keepaspectratio]{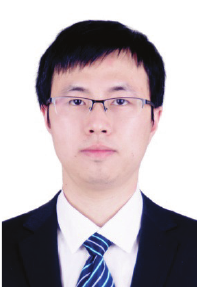}}]{Wenchi Cheng} (M'14-SM'18) received the B.S. and Ph.D. degrees in telecommunication engineering from Xidian University, Xian, China, in 2008 and 2013, respectively, where he is a Full Professor. He was a Visiting Scholar with Networking and Information Systems Laboratory, Department of Electrical and Computer Engineering, Texas A\&M University, College Station, TX, USA, from 2010 to 2011. His current research interests include B5G/6G wireless networks, emergency wireless communications, and orbital-angular-momentum based wireless communications. He has published more than 100 international journal and conference papers in IEEE Journal on Selected Areas in Communications, IEEE Magazines, IEEE Transactions, IEEE INFOCOM, GLOBECOM, and ICC, etc. He received URSI Young Scientist Award (2019), the Young Elite Scientist Award of CAST, the Best Dissertation of China Institute of Communications, the Best Paper Award for IEEE ICCC 2018, the Best Paper Award for IEEE WCSP 2019, and the Best Paper Nomination for IEEE GLOBECOM 2014. He has served or serving as the Associate Editor for IEEE Systems Journal, IEEE Communications Letters, IEEE Wireless Communications Letter, the IoT Session Chair for IEEE 5G Roadmap, the Wireless Communications Symposium Co-Chair for IEEE ICC 2022 and IEEE GLOBECOM 2020, the Publicity Chair for IEEE ICC 2019, the Next Generation Networks Symposium Chair for IEEE ICCC 2019, the Workshop Chair for IEEE ICC 2019/IEEE GLOBECOM 2019/INFOCOM 2020 Workshop on Intelligent Wireless Emergency Communications Networks.
\end{IEEEbiography}
\begin{IEEEbiography}
[{\includegraphics[width=1in,height=1.25in,clip,keepaspectratio]{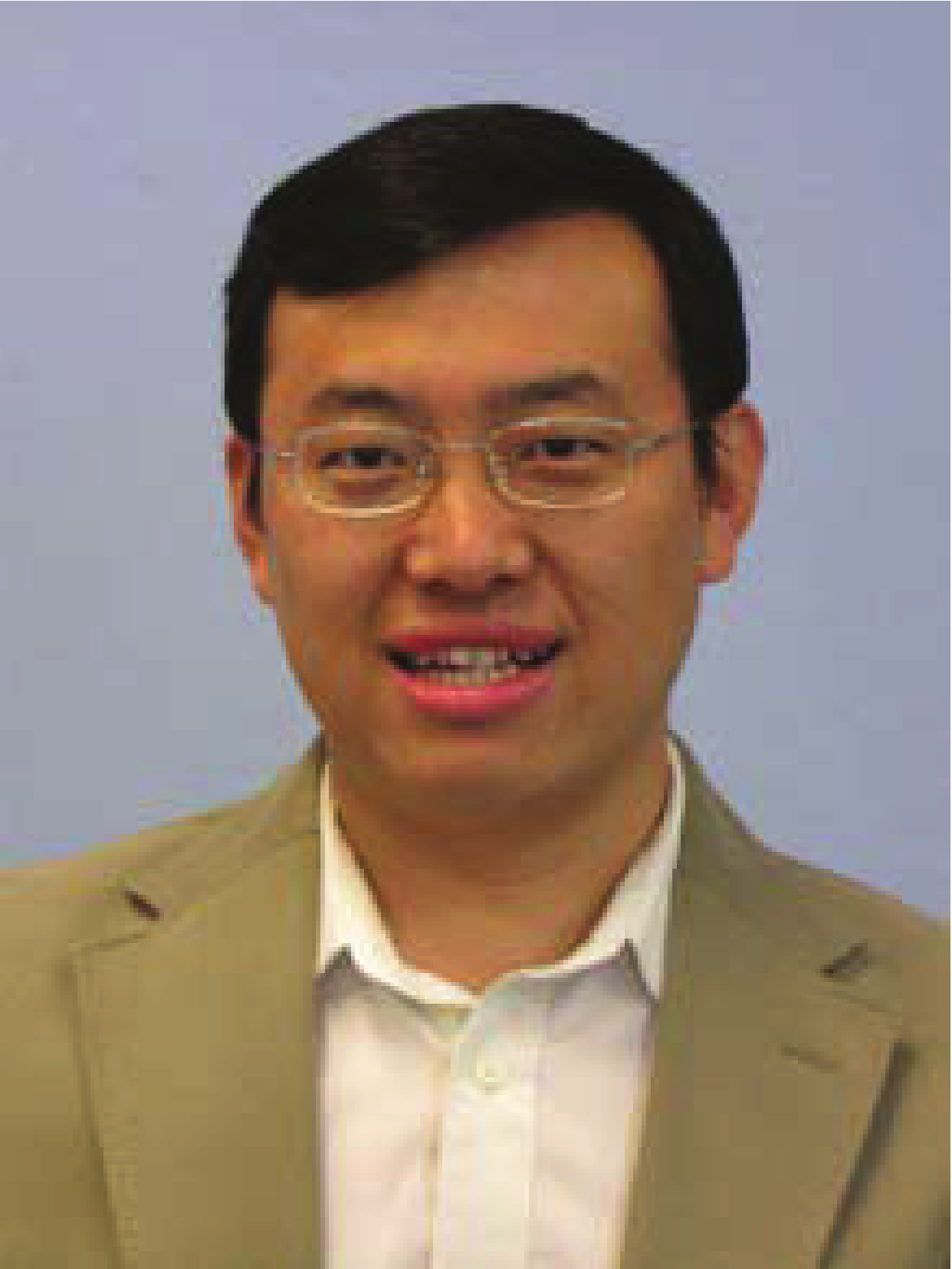}}]{Wei Zhang} (S'01-M'06-SM'11-F'15) received the Ph.D. degree from The Chinese University of Hong Kong in 2005. Currently, he is a Professor at the School of Electrical Engineering and Telecommunications, the University of New South Wales, Sydney, Australia. His current research interests include UAV communications, 5G and beyond. He received 6 best paper awards from IEEE conferences and ComSoc technical committees. He was elevated to Fellow of the IEEE in 2015 and was an IEEE ComSoc Distinguished Lecturer in 2016-2017.

Within the IEEE ComSoc, he has taken many leadership positions including Member-at-Large on the Board of Governors (2018-2020), Chair of Wireless Communications Technical Committee (2019-2020), Vice Director of Asia Pacific Board (2016-2021), Editor-in-Chief of IEEE Wireless Communications Letters (2016-2019), Technical Program Committee Chair of APCC 2017 and ICCC 2019, Award Committee Chair of Asia Pacific Board and Award Committee Chair of Technical Committee on Cognitive Networks. He was recently elected as Vice President of IEEE Communications Society (2022-2023).

In addition, he has served as a member in various ComSoc boards/standing committees, including Journals Board, Technical Committee Recertification Committee, Finance Standing Committee, Information Technology Committee, Steering Committee of IEEE Transactions on Green Communications and Networking and Steering Committee of IEEE Networking Letters. Currently, he serves as an Area Editor of the IEEE Transactions on Wireless Communications and the Editor-in-Chief of Journal of Communications and Information Networks. Previously, he served as Editor of IEEE Transactions on Communications, IEEE Transactions on Wireless Communications, IEEE Transactions on Cognitive Communications and Networking, and IEEE Journal on Selected Areas in Communications C Cognitive Radio Series.
\end{IEEEbiography}
\vspace{-51 mm}
\begin{IEEEbiography}[{\includegraphics[width=1in,height=1.25in,clip,keepaspectratio]{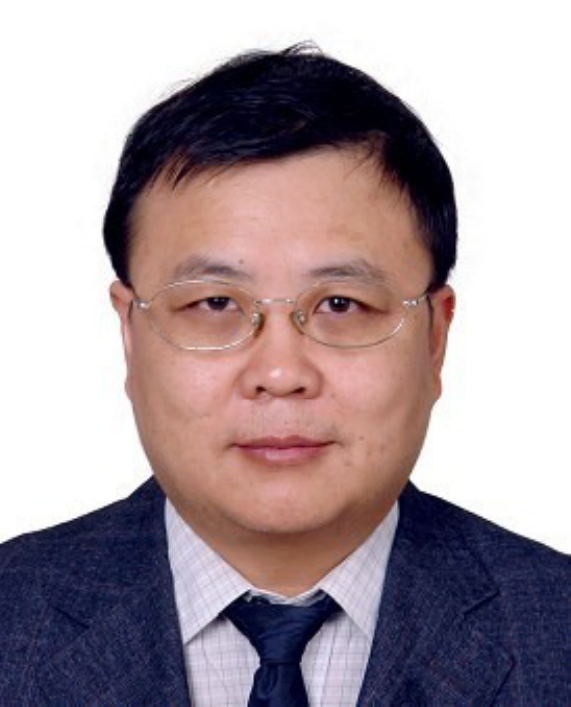}}]{Hailin Zhang} (M'97) received B.S. and M.S. degrees from Northwestern Polytechnic University, Xi'an, China, in 1985 and 1988 respectively, and the Ph.D. from Xidian University, Xi'an, China, in 1991. In 1991, he joined School of Telecommunications Engineering, Xidian University, where he is a senior Professor and the Dean of this school. He is also currently the Director of Key Laboratory in Wireless Communications Sponsored by China Ministry of Information Technology, a key member of State Key Laboratory of Integrated Services Networks, one of the state government specially compensated scientists and engineers, a field leader in Telecommunications and Information Systems in Xidian University, an Associate Director of National 111 Project. Dr. Zhang's current research interests include key transmission technologies and standards on broadband wireless communications for B5G/6G wireless access systems. He has published more than 200 papers in journals and conferences.
\end{IEEEbiography}

\end{document}